\let\includefigures=\iftrue
%
\let\useblackboard=\iftrue
%
%
\newfam\black
\input harvmac
\noblackbox

\useblackboard
\message{If you do not have msbm (blackboard bold) fonts,}
\message{change the option at the top of the tex file.}
\font\blackboard=msbm10 scaled \magstep1
\font\blackboards=msbm7
\font\blackboardss=msbm5
\textfont\black=\blackboard
\scriptfont\black=\blackboards
\scriptscriptfont\black=\blackboardss

\else

\fi
\def\boxit#1{\vbox{\hrule\hbox{\vrule\kern8pt
\vbox{\hbox{\kern8pt}\hbox{\vbox{#1}}\hbox{\kern8pt}}
\kern8pt\vrule}\hrule}}
\def\mathboxit#1{\vbox{\hrule\hbox{\vrule\kern8pt\vbox{\kern8pt
\hbox{$\displaystyle #1$}\kern8pt}\kern8pt\vrule}\hrule}}

\def\subsubsection#1{\bigskip\noindent
{\it #1}}

\def\yboxit#1#2{\vbox{\hrule height #1 \hbox{\vrule width #1
\vbox{#2}\vrule width #1 }\hrule height #1 }}
\def\fillbox#1{\hbox to #1{\vbox to #1{\vfil}\hfil}}
\def\ybox{{\lower 1.3pt \yboxit{0.4pt}{\fillbox{8pt}}\hskip-0.2pt}}
%


\def\bphi{{\bar \phi}}


\def\bz{{\bar z}}

\def\l{\left}

\def\comments#1{}

\def\p{\partial}

\def\eps{\epsilon}
\def\half{{1\over 2}}

\def\CC{{\cal C}}

\def\CF{{\cal F}}

\def\CO{{\cal O}}

\def\a{\alpha}

\def\II{\relax{I\kern-.10em I}}

\font\cmss=cmss10 \font\cmsss=cmss10 at 7pt
\def\IZ{\relax\ifmmode\mathchoice
{\hbox{\cmss Z\kern-.4em Z}}{\hbox{\cmss Z\kern-.4em Z}}
{\lower.9pt\hbox{\cmsss Z\kern-.4em Z}}
{\lower1.2pt\hbox{\cmsss Z\kern-.4em Z}}
\else{\cmss Z\kern-.4emZ}\fi}
\def\IR{\relax{\rm I\kern-.18em R}}
\def\IZ{\relax\ifmmode\mathchoice
{\hbox{\cmss Z\kern-.4em Z}}{\hbox{\cmss Z\kern-.4em Z}}
{\lower.9pt\hbox{\cmsss Z\kern-.4em Z}} {\lower1.2pt\hbox{\cmsss
Z\kern-.4em Z}}\else{\cmss Z\kern-.4em Z}\fi}
\def\IB{\relax{\rm I\kern-.18em B}}
\def\IC{{\relax\hbox{$\inbar\kern-.3em{\rm C}$}}}
\def\ID{\relax{\rm I\kern-.18em D}}
\def\IE{\relax{\rm I\kern-.18em E}}
\def\IF{\relax{\rm I\kern-.18em F}}
\def\IG{\relax\hbox{$\inbar\kern-.3em{\rm G}$}}
\def\IGa{\relax\hbox{${\rm I}\kern-.18em\Gamma$}}
\def\IH{\relax{\rm I\kern-.18em H}}
\def\II{\relax{\rm I\kern-.18em I}}
\def\IK{\relax{\rm I\kern-.18em K}}
\def\IP{\relax{\rm I\kern-.18em P}}

\def\inbar{\,\vrule height1.5ex width.4pt depth0pt}

\def\p{\partial}

\font\cmss=cmss10 
\def\IR{\relax{\rm I\kern-.18em R}}

\def\lp10{\ell_p^{10}}
\def\lp11{\ell_p^{11}}
\def\R11{R_{11}}

\def\frac#1#2{{#1 \over #2}}


\def\l{\left}

\def\comments#1{}

\def\p{\partial}

\def\eps{\epsilon}
\def\half{{1\over 2}}

\def\CC{{\cal C}}

\def\CF{{\cal F}}

\def\CO{{\cal O}}


\def\cf{{\it c.f.}}

\def\M4{M_{Pl,4}}

\def\k11{\kappa_{11}}
\def\l11{\ell_{11}}
\def\tl11{\tilde{\ell}_{11}}

\def\m11{M_{11}}
\def\tm11{\tilde{M}_{11}}




\lref\TseytlinCD{
  A.~A.~Tseytlin,
  ``On 'rolling moduli' solutions in string cosmology,''
  Phys.\ Lett.\ B {\bf 334}, 315 (1994)
  [arXiv:hep-th/9404191].
}

\lref\SuyamaHW{
  T.~Suyama,
  ``Tachyons in compact spaces,''
  JHEP {\bf 0505}, 065 (2005)
  [arXiv:hep-th/0503073].
}

\lref\BanksQS{
 T.~Banks and E.~J.~Martinec,
 ``The Renormalization Group And String Field Theory,''
 Nucl.\ Phys.\ B {\bf 294}, 733 (1987).
}

\lref\VafaUE{
 C.~Vafa,
 ``C Theorem And The Topology Of 2-D Qfts,''
 Phys.\ Lett.\ B {\bf 212}, 28 (1988).
}

\lref\SuyamaWD{
  T.~Suyama,
  ``Closed String Tachyon Condensation in Supercritical Strings and RG Flows,''
  arXiv:hep-th/0510174.
}

\lref\HellermanQA{
  S.~Hellerman and X.~Liu,
  ``Dynamical dimension change in supercritical string theory,''
  arXiv:hep-th/0409071.
}

\lref\DasDA{
  S.~R.~Das, A.~Dhar and S.~R.~Wadia,
  ``Critical Behavior In Two-Dimensional Quantum Gravity And Equations Of
  Motion Of The String,''
  Mod.\ Phys.\ Lett.\ A {\bf 5}, 799 (1990).
}

\lref\DavidVM{
  J.~R.~David, M.~Gutperle, M.~Headrick and S.~Minwalla,
  ``Closed string tachyon condensation on twisted circles,''
  JHEP {\bf 0202}, 041 (2002)
  [arXiv:hep-th/0111212].
}
\lref\HeadrickHZ{
  M.~Headrick, S.~Minwalla and T.~Takayanagi,
  ``Closed string tachyon condensation: An overview,''
  Class.\ Quant.\ Grav.\  {\bf 21}, S1539 (2004)
  [arXiv:hep-th/0405064].
}
\lref\FriedanXQ{
  D.~Friedan, Z.~a.~Qiu and S.~H.~Shenker,
  ``Conformal Invariance, Unitarity And Two-Dimensional Critical Exponents,''
  Phys.\ Rev.\ Lett.\  {\bf 52}, 1575 (1984).
}
\lref\ZamolodchikovDB{
  A.~B.~Zamolodchikov,
  ``Conformal Symmetry And Multicritical Points In Two-Dimensional Quantum
  Field Theory. (In Russian),''
  Sov.\ J.\ Nucl.\ Phys.\  {\bf 44}, 529 (1986)
  [Yad.\ Fiz.\  {\bf 44}, 821 (1986)].
}
\lref\MukherjiTB{
  S.~Mukherji and A.~Sen,
  ``Some all order classical solutions in nonpolynomial closed string field
  theory,''
  Nucl.\ Phys.\ B {\bf 363}, 639 (1991).
}
\lref\DabholkarKY{
  A.~Dabholkar, A.~Iqubal and J.~Raeymaekers,
  ``Off-shell interactions for closed-string tachyons,''
  JHEP {\bf 0405}, 051 (2004)
  [arXiv:hep-th/0403238].
}
\lref\DineCA{
  M.~Dine, E.~Gorbatov, I.~R.~Klebanov and M.~Krasnitz,
  ``Closed string tachyons and their implications for non-supersymmetric
  strings,''
  JHEP {\bf 0407}, 034 (2004)
  [arXiv:hep-th/0303076].
}
\lref\FradkinYS{
 E.~S.~Fradkin and A.~A.~Tseytlin,
 ``Quantum String Theory Effective Action,''
 Nucl.\ Phys.\ B {\bf 261}, 1 (1985).
}
\lref\FradkinPQ{
 E.~S.~Fradkin and A.~A.~Tseytlin,
 ``Effective Field Theory From Quantized Strings,''
 Phys.\ Lett.\ B {\bf 158}, 316 (1985).
}
\lref\DasCZ{
 S.~R.~Das and B.~Sathiapalan,
 ``String Propagation In A Tachyon Background,''
 Phys.\ Rev.\ Lett.\  {\bf 56}, 2664 (1986).
}
\lref\JainJV{
 S.~Jain, G.~Mandal and S.~R.~Wadia,
 ``Perturbatively Renormalized Vertex Operator, Highest Weight Representations
 Of Virasoro Algebra And String Dynamics In Curved Space,''
 Phys.\ Rev.\ D {\bf 35}, 3116 (1987).
}
\lref\CooperVG{
 A.~R.~Cooper, L.~Susskind and L.~Thorlacius,
 ``Two-dimensional quantum cosmology,''
 Nucl.\ Phys.\ B {\bf 363}, 132 (1991).
}
\lref\FriedanXQ{
  D.~Friedan, Z.~a.~Qiu and S.~H.~Shenker,
  ``Conformal Invariance, Unitarity And Two-Dimensional Critical Exponents,''
  Phys.\ Rev.\ Lett.\  {\bf 52}, 1575 (1984).
}

\lref\FriedanII{
  D.~Friedan,
  ``On Two-Dimensional Conformal Invariance And The Field Theory Of String,''
  Phys.\ Lett.\ B {\bf 162}, 102 (1985).
}
\lref\MartinecSP{
  E.~J.~Martinec,
  ``The Space Of 2-D Quantum Field Theories,''
  from {\it The Interface of Mathematics and Particle Physics}, D.G. Quillen\ {\it et. al.},
  eds., (Oxford Univ. Press, 1990).
}

\lref\SenNU{
A.~Sen,
``Rolling tachyon,''
JHEP {\bf 0204}, 048 (2002)
[arXiv:hep-th/0203211].
}
\lref\SenXM{
A.~Sen,
``Universality of the tachyon potential,''
JHEP {\bf 9912}, 027 (1999)
[arXiv:hep-th/9911116].
}

\lref\PolyakovRD{
A.~M.~Polyakov,
``Quantum Geometry Of Bosonic Strings,''
Phys.\ Lett.\ B {\bf 103}, 207 (1981);
A.~M.~Polyakov,
``Quantum Geometry Of Fermionic Strings,''
Phys.\ Lett.\ B {\bf 103}, 211 (1981).
}
\lref\PolyakovTP{
A.~M.~Polyakov,
``A Few projects in string theory,''
arXiv:hep-th/9304146.
}
\lref\KlebanovMS{
I.~R.~Klebanov, I.~I.~Kogan and A.~M.~Polyakov,
``Gravitational dressing of renormalization group,''
Phys.\ Rev.\ Lett.\  {\bf 71}, 3243 (1993)
[arXiv:hep-th/9309106].
}

\lref\SchmidhuberBV{
C.~Schmidhuber and A.~A.~Tseytlin,
``On string cosmology and the RG flow in 2-d field theory,''
Nucl.\ Phys.\ B {\bf 426}, 187 (1994)
[arXiv:hep-th/9404180].
}

\lref\GinspargIS{
P.~H.~Ginsparg and G.~W.~Moore,
``Lectures on 2-D gravity and 2-D string theory,''
arXiv:hep-th/9304011.
}
\lref\DavidHJ{
F.~David,
``Conformal Field Theories Coupled To 2-D Gravity In The Conformal Gauge,''
Mod.\ Phys.\ Lett.\ A {\bf 3}, 1651 (1988).
J.~Distler and H.~Kawai,
``Conformal Field Theory And 2-D Quantum Gravity Or Who's Afraid Of Joseph
Liouville?,''
Nucl.\ Phys.\ B {\bf 321}, 509 (1989).
}
\lref\KnizhnikAK{
V.~G.~Knizhnik, A.~M.~Polyakov and A.~B.~Zamolodchikov,
``Fractal Structure Of 2d-Quantum Gravity,''
Mod.\ Phys.\ Lett.\ A {\bf 3}, 819 (1988).
}
\lref\SeibergEB{
N.~Seiberg,
``Notes On Quantum Liouville Theory And Quantum Gravity,''
Prog.\ Theor.\ Phys.\ Suppl.\  {\bf 102}, 319 (1990).
}
\lref\martinecrev{E.~Martinec, "An introduction to 2D gravity and solvable string models"}

\lref\DharAI{
A.~Dhar and S.~R.~Wadia,
``Noncritical strings, RG flows and holography,''
Nucl.\ Phys.\ B {\bf 590}, 261 (2000)
[arXiv:hep-th/0006043].
}
\lref\ZamolodchikovJB{
  A.~Zamolodchikov,
  ``Perturbed conformal field theory on fluctuating sphere,''
  arXiv:hep-th/0508044.
}
\lref\PolchinskiMH{
  J.~Polchinski,
  ``Remarks On The Liouville Field Theory,''
UTTG-19-90
{\it Presented at Strings '90 Conf., College Station, TX, Mar 12-17, 1990}
}
\lref\GinspargIS{
  P.~H.~Ginsparg and G.~W.~Moore,
  ``Lectures on 2-D gravity and 2-D string theory,''
  arXiv:hep-th/9304011.
}
\lref\KutasovFG{
  D.~Kutasov, K.~Okuyama, J.~w.~Park, N.~Seiberg and D.~Shih,
  ``Annulus amplitudes and ZZ branes in minimal string theory,''
  JHEP {\bf 0408}, 026 (2004)
  [arXiv:hep-th/0406030].
}

\lref\GiddingsDE{
S.~B.~Giddings, J.~A.~Harvey, J.~G.~Polchinski, S.~H.~Shenker and A.~Strominger,
``Hairy black holes in string theory,''
Phys.\ Rev.\ D {\bf 50}, 6422 (1994)
[arXiv:hep-th/9309152].
}

\lref\WittenQY{
E.~Witten,
``On background independent open string field theory,''
Phys.\ Rev.\ D {\bf 46}, 5467 (1992)
[arXiv:hep-th/9208027].
}
\lref\WittenCR{
E.~Witten,
``Some computations in background independent off-shell string theory,''
Phys.\ Rev.\ D {\bf 47}, 3405 (1993)
[arXiv:hep-th/9210065].
}
\lref\ShatashviliKK{
S.~L.~Shatashvili,
``Comment on the background independent open string theory,''
Phys.\ Lett.\ B {\bf 311}, 83 (1993)
[arXiv:hep-th/9303143].
}
\lref\ShatashviliPS{
S.~L.~Shatashvili,
``On the problems with background independence in string theory,''
Alg.\ Anal.\  {\bf 6}, 215 (1994)
[arXiv:hep-th/9311177].
}
\lref\GerasimovZP{
A.~A.~Gerasimov and S.~L.~Shatashvili,
``On exact tachyon potential in open string field theory,''
JHEP {\bf 0010}, 034 (2000)
[arXiv:hep-th/0009103].
}
\lref\GerasimovGA{
A.~A.~Gerasimov and S.~L.~Shatashvili,
``Stringy Higgs mechanism and the fate of open strings,''
JHEP {\bf 0101}, 019 (2001)
[arXiv:hep-th/0011009].
}
\lref\KutasovQP{
D.~Kutasov, M.~Marino and G.~W.~Moore,
``Some exact results on tachyon condensation in string field theory,''
JHEP {\bf 0010}, 045 (2000)
[arXiv:hep-th/0009148].
}
\lref\KutasovAQ{
D.~Kutasov, M.~Marino and G.~W.~Moore,
``Remarks on tachyon condensation in superstring field theory,''
arXiv:hep-th/0010108.
}

\lref\HarveyNA{
J.~A.~Harvey, D.~Kutasov and E.~J.~Martinec,
``On the relevance of tachyons,''
arXiv:hep-th/0003101.
}

\lref\FriedanCS{
D.~Friedan and S.~H.~Shenker,
``The Integrable Analytic Geometry Of Quantum String,''
Phys.\ Lett.\ B {\bf 175}, 287 (1986).
}
\lref\FriedanUA{
D.~Friedan and S.~H.~Shenker,
``The Analytic Geometry Of Two-Dimensional Conformal Field Theory,''
Nucl.\ Phys.\ B {\bf 281}, 509 (1987).
}
\lref\friedan{D.~Friedan, SFT in space of 2d QFTs}
\lref\martinectwod{E.~J.~Martinec, SFT and 2d QFTs}

\lref\FriedanAA{
D.~Friedan,
``A tentative theory of large distance physics,''
JHEP {\bf 0310}, 063 (2003)
[arXiv:hep-th/0204131].
}


\lref\FriedanJM{
  D.~H.~Friedan,
  ``Nonlinear Models In Two + Epsilon Dimensions,''
  Annals Phys.\  {\bf 163}, 318 (1985).
}
\lref\AlvarezGaumeHN{
  L.~Alvarez-Gaume, D.~Z.~Freedman and S.~Mukhi,
  ``The Background Field Method And The Ultraviolet Structure Of The
  Supersymmetric Nonlinear Sigma Model,''
  Annals Phys.\  {\bf 134}, 85 (1981).
}
\lref\CallanIA{
  C.~G.~.~Callan, E.~J.~Martinec, M.~J.~Perry and D.~Friedan,
  ``Strings In Background Fields,''
  Nucl.\ Phys.\ B {\bf 262}, 593 (1985).
}
\lref\LovelaceKR{
  C.~Lovelace,
  ``Stability Of String Vacua. 1. A New Picture Of The Renormalization Group,''
  Nucl.\ Phys.\ B {\bf 273}, 413 (1986).
}
\lref\CallanJB{
  C.~G.~.~Callan, I.~R.~Klebanov and M.~J.~Perry,
  ``String Theory Effective Actions,''
  Nucl.\ Phys.\ B {\bf 278}, 78 (1986).
}
\lref\TseytlinTT{
  A.~A.~Tseytlin,
  ``Conformal Anomaly In Two-Dimensional Sigma Model On Curved Background And
  Strings,''
  Phys.\ Lett.\  {\bf 178B}, 34 (1986).
}
\lref\TseytlinWS{
  A.~A.~Tseytlin,
  ``Sigma Model Weyl Invariance Conditions And String Equations Of Motion,''
  Nucl.\ Phys.\ B {\bf 294}, 383 (1987).
}

\lref\BelavinVU{
  A.~A.~Belavin, A.~M.~Polyakov and A.~B.~Zamolodchikov,
  ``Infinite Conformal Symmetry In Two-Dimensional Quantum Field Theory,''
  Nucl.\ Phys.\ B {\bf 241}, 333 (1984).
}

\lref\KosterlitzSM{
  J.~M.~Kosterlitz,
  ``The Critical Properties Of The Two-Dimensional XY Model,''
  J.\ Phys.\ C {\bf 7}, 1046 (1974).
}
\lref\AmitAB{
  D.~J.~Amit, Y.~Y.~Goldschmidt and G.~Grinstein,
  ``Renormalization Group Analysis Of The Phase Transition In The 2 D Coulomb
  Gas, Sine-Gordon Theory And XY Model,''
  J.\ Phys.\ A {\bf 13}, 585 (1980).
}
\lref\LudwigRK{
  A.~W.~W.~Ludwig,
  ``Critical Behavior Of The Two-Dimensional Random Q State Potts Model By
  Expansion In (Q-2),''
  Nucl.\ Phys.\ B {\bf 285}, 97 (1987).
}
\lref\LudwigGS{
  A.~W.~W.~Ludwig and J.~L.~Cardy,
  ``Perturbative Evaluation Of The Conformal Anomaly At New Critical Points
  With Applications To Random Systems,''
  Nucl.\ Phys.\ B {\bf 285}, 687 (1987).
}
\lref\ZamolodchikovTI{
  A.~B.~Zamolodchikov,
  ``Renormalization Group And Perturbation Theory Near Fixed Points In
  Two-Dimensional Field Theory,''
  Sov.\ J.\ Nucl.\ Phys.\  {\bf 46}, 1090 (1987)
  [Yad.\ Fiz.\  {\bf 46}, 1819 (1987)].
}

\lref\ZamolodchikovBK{
  A.~B.~Zamolodchikov,
  ``Two Point Correlation Function In Scaling Lee-Yang Model,''
  Nucl.\ Phys.\ B {\bf 348}, 619 (1991).
}
\lref\BelavinPU{
  A.~A.~Belavin, V.~A.~Belavin, A.~V.~Litvinov, Y.~P.~Pugai and A.~B.~Zamolodchikov,
  ``On correlation functions in the perturbed minimal models M(2,2n+1),''
  Nucl.\ Phys.\ B {\bf 676}, 587 (2004)
  [arXiv:hep-th/0309137].
}
\lref\LeafHerrmannDB{
  W.~A.~Leaf-Herrmann,
  ``Perturbation Theory Near N=2 Superconformal Fixed Points In Two-Dimensional
  Field Theory,''
  Nucl.\ Phys.\ B {\bf 348}, 525 (1991).
}
\lref\KastorEF{
  D.~A.~Kastor, E.~J.~Martinec and S.~H.~Shenker,
  ``RG Flow In N=1 Discrete Series,''
  Nucl.\ Phys.\ B {\bf 316}, 590 (1989).
}
\lref\SenMZ{
  A.~Sen,
  ``Nontrivial Renormalization Group Fixed Points And Solutions Of String Field
  Theory Equations Of Motion,''
  Phys.\ Lett.\ B {\bf 252}, 566 (1990).
}
\lref\DotsenkoSY{
  V.~Dotsenko, M.~Picco and P.~Pujol,
  ``Renormalization group calculation of correlation functions for the 2-d
  random bond Ising and Potts models,''
  Nucl.\ Phys.\ B {\bf 455}, 701 (1995)
  [arXiv:hep-th/9501017].
}

\lref\AffleckTK{
  I.~Affleck and A.~W.~W.~Ludwig,
  ``Universal noninteger 'ground state degeneracy' in critical quantum
  systems,''
  Phys.\ Rev.\ Lett.\  {\bf 67}, 161 (1991).
}
\lref\HarveyGQ{
  J.~A.~Harvey, S.~Kachru, G.~W.~Moore and E.~Silverstein,
  ``Tension is dimension,''
  JHEP {\bf 0003}, 001 (2000)
  [arXiv:hep-th/9909072].
}

\lref\AdamsSV{
  A.~Adams, J.~Polchinski and E.~Silverstein,
  ``Don't panic! Closed string tachyons in ALE space-times,''
  JHEP {\bf 0110}, 029 (2001)
  [arXiv:hep-th/0108075].
}
\lref\GutperleKI{
  M.~Gutperle, M.~Headrick, S.~Minwalla and V.~Schomerus,
  ``Space-time energy decreases under world-sheet RG flow,''
  JHEP {\bf 0301}, 073 (2003)
  [arXiv:hep-th/0211063].
}

\lref\ZamolodchikovGT{
  A.~B.~Zamolodchikov,
  ``'Irreversibility' Of The Flux Of The Renormalization Group In A 2-D Field
  Theory,''
  JETP Lett.\  {\bf 43}, 730 (1986)
  [Pisma Zh.\ Eksp.\ Teor.\ Fiz.\  {\bf 43}, 565 (1986)].
}
\lref\deAlwisPS{
  S.~P.~de Alwis,
  ``The C Theorem, The Dilaton And The Effective Action In String Theory,''
  Phys.\ Lett.\ B {\bf 217}, 467 (1989).
}

\lref\KutasovXB{
  D.~Kutasov,
  ``Geometry On The Space Of Conformal Field Theories And Contact Terms,''
  Phys.\ Lett.\ B {\bf 220}, 153 (1989).
}

\lref\DabholkarWN{
  A.~Dabholkar and C.~Vafa,
  ``tt* geometry and closed string tachyon potential,''
  JHEP {\bf 0202}, 008 (2002)
  [arXiv:hep-th/0111155].
}
\lref\CecottiME{
  S.~Cecotti and C.~Vafa,
  ``Topological antitopological fusion,''
  Nucl.\ Phys.\ B {\bf 367}, 359 (1991).
}
\lref\KosteleckyMU{
  V.~A.~Kostelecky, M.~Perry and R.~Potting,
  ``Off-shell structure of the string sigma model,''
  Phys.\ Rev.\ Lett.\  {\bf 84}, 4541 (2000)
  [arXiv:hep-th/9912243].
}

\lref\AffleckTK{
  I.~Affleck and A.~W.~W.~Ludwig,
  ``Universal noninteger 'ground state degeneracy' in critical quantum
  systems,''
  Phys.\ Rev.\ Lett.\  {\bf 67}, 161 (1991).
}
\lref\FriedanYC{
  D.~Friedan and A.~Konechny,
  ``On the boundary entropy of one-dimensional quantum systems at low
  temperature,''
  Phys.\ Rev.\ Lett.\  {\bf 93}, 030402 (2004)
  [arXiv:hep-th/0312197].
}

\lref\GutperleAI{
  M.~Gutperle and A.~Strominger,
  ``Spacelike branes,''
  JHEP {\bf 0204}, 018 (2002)
  [arXiv:hep-th/0202210].
}

\lref\BalasubramanianSN{
  V.~Balasubramanian, P.~Kraus and A.~E.~Lawrence,
  ``Bulk vs. boundary dynamics in anti-de Sitter spacetime,''
  Phys.\ Rev.\ D {\bf 59}, 046003 (1999)
  [arXiv:hep-th/9805171].
}
\lref\BalasubramanianDE{
  V.~Balasubramanian, P.~Kraus, A.~E.~Lawrence and S.~P.~Trivedi,
  ``Holographic probes of anti-de Sitter space-times,''
  Phys.\ Rev.\ D {\bf 59}, 104021 (1999)
  [arXiv:hep-th/9808017].
}

\lref\BalasubramanianJD{
  V.~Balasubramanian and P.~Kraus,
  ``Spacetime and the holographic renormalization group,''
  Phys.\ Rev.\ Lett.\  {\bf 83}, 3605 (1999)
  [arXiv:hep-th/9903190].
}
\lref\deBoerXF{
  J.~de Boer, E.~P.~Verlinde and H.~L.~Verlinde,
  ``On the holographic renormalization group,''
  JHEP {\bf 0008}, 003 (2000)
  [arXiv:hep-th/9912012].
}

\lref\polyakovbook{A.~M.~Polyakov, {\it Gauge Fields and Strings}, Harwood, UK (1987).}
\lref\cardybook{J.~Cardy, {\it Scaling and Renormalization in Statistical Physics}, Cambridge
Univ. Press, UK, 1996.}
\lref\PolchinskiRR{
  J.~Polchinski,
  {\it String theory. Vol. 2: Superstring theory and beyond,}, (Cambridge Univ. Press, UK, 1998).
}
\lref\CardyDA{
  J.~L.~Cardy,
  ``Conformal Invariance And Statistical Mechanics,'' 1988 Les Houches lectures
}

\Title{\vbox{\baselineskip12pt\hbox{hep-th/0510126}
\hbox{MIT-CTP-3693}
\hbox{BRX TH-558}}}
{\vbox{
\centerline{On Closed String Tachyon Dynamics}
}}
\smallskip
\centerline{Daniel Z. Freedman${}^{1,2}$, Matthew Headrick${}^2$, and Albion Lawrence${}^3$}
\smallskip
\centerline{${}^1$ {\it Department of Mathematics, Massachusetts Institute of Technology}}
\centerline{{\it Cambridge MA 02139, USA}}
\smallskip
\centerline{${}^2$ {\it Center for Theoretical Physics, Massachusetts Institute of Technology}}
\centerline{{\it Cambridge MA 02139, USA}}
\smallskip
\centerline{{${}^3$ {\it Martin Fisher School of Physics, Brandeis 
University}}}
\centerline{{\it MS 057, PO Box 549110, Waltham MA 02454, USA}}

\bigskip
\bigskip
\noindent

We study the condensation of closed string tachyons as a time-dependent process. In particular, we study tachyons whose wave functions are either space-filling or localized in a compact space, and whose masses are small in string units; our analysis is otherwise general and does not depend on any specific model. Using world-sheet methods, we calculate the equations of motion for the coupled tachyon-dilaton system, and show that the tachyon follows geodesic motion with respect to the Zamolodchikov metric, subject to a force proportional to its beta function and friction proportional to the time derivative of the dilaton. We study the relationship between world-sheet RG flow and the solutions to our equations, finding a close relationship in the case that the spatial theory is supercritical and the dilaton has a negative time derivative.

\Date{October 2005}
\listtoc
\writetoc

\newsec{Introduction}

Two outstanding problems in string theory are how to understand the
configuration space of the theory, and how to understand time-dependent
backgrounds.  These problems are inseparable. The statement that
two classical vacua are connected in some configuration space is
meaningful if there is some dynamical process (or domain wall) which
interpolates between them.

In this paper we report modest progress towards understanding these issues in
classical closed string theory.
To do so, we will focus on a particular class of time-dependent backgrounds: 
the decay of unstable vacua via closed string
tachyon condensation.  Such unstable vacua
correspond to two-dimensional conformal field theories (CFTs) with relevant operators.
The standard lore, based on the study of many examples in open and closed string theory, 
is that perturbing the background by a tachyon condensate is somehow equivalent to perturbing the world-sheet  CFT by the corresponding relevant operator; and that the 
endpoint of tachyon condensation is the endpoint of the renormalization group flow
of this perturbed CFT. This is in accord with a philosophy espoused by some
since the mid-1980s, that the (classical) configuration space of string theory is the space
of all two-dimensional quantum field theories ({\it c.f.}\ \refs{\FriedanII,\MartinecSP,\BanksQS,\VafaUE}).\foot{In general even this is too simple.  In closed string theory
we must be able to take into account D-branes and Ramond-Ramond backgrounds.
Furthermore, it is not clear how to deal with perturbations by massive fields.
We will avoid all such backgrounds in this work.}

This picture cannot be the complete story.
To begin with, the RG flows are governed by equations that are first order in
derivatives with respect to a world-sheet cutoff, whereas the
spacetime equations are second or higher order.
Secondly, except in cases where the closed string tachyon mode is localized in a non-compact space, the central charge of the perturbed world-sheet field theory must strictly decrease along the
renormalization group flow \refs{\ZamolodchikovGT,\ZamolodchikovTI}.  However, under time evolution the central charge of the theory cannot change, since it must remain critical.

We therefore wish to study the spacetime dynamics of tachyons more directly.
Inspired by \refs{\DavidHJ,\SchmidhuberBV,\PolyakovTP},\foot{In particular, this
program was carried out for sigma models without tachyons in \refs{\SchmidhuberBV,\TseytlinCD}.}
we consider the following backgrounds.
Begin with a CFT $\CC$ with a set of relevant and marginal operators $\CO_a$. 
Perturb the theory by turning on couplings $u^a$ to these operators:\foot{For convenience, 
we use a notation that assumes that $\CC$ has a Lagrangian description. If not, the 
perturbed theory may be defined via the correlation functions 
$ \langle \CO_1(x_1)\ldots \CO_n(x_n)\rangle_{\bar S_u} \equiv \langle \CO_1(x_1)
\ldots \CO_n(x_n) e^{-\delta \bar S}\rangle_\CC$.}
\eqn\perturbedaction{
\bar S_u + S_\Phi = S_\CC + \int d^2 z\left(\eps^{\Delta_a - 2} u^a \CO_a + {1\over4\pi\alpha'}\Phi R^{(2)}\right).
}
Here $\Delta_a$ is the dimension of $\CO_a$, $\eps$ is a cutoff with dimensions of length,
and $u^a$ is a dimensionless coupling. We have also included a dilaton $\Phi$ 
(which for the moment is constant) coupling to the world-sheet curvature $R^{(2)}$. 
The theory \perturbedaction\ will not in general be conformal, and the couplings 
$u^a,\Phi$ will have beta functions $\bar\beta^a(u),\bar\beta^\Phi(u)$ which we 
assume are known. Note that at a fixed point $\bar\beta^\Phi$ is equal to one-sixth 
the central charge.

Now we couple this theory to a scalar field representing the time direction
in the target space:
\eqn\gravdressed{
	S_{u(\phi)} + S_{\Phi(\phi)} = S_\CC + \int d^2 z \left[ - {1\over2\pi\alpha'}\p\phi\bar\p\phi + \eps^{\Delta_a - 2} u^a(\phi) \CO_a + {1\over4\pi\alpha'}\Phi(\phi) R^{(2)} \right].
}
The fields $u^a$, as well as the dilaton, are now time-dependent. The target space 
equations of motion are the conditions that the theory perturbed by $S_{u(\phi)}$ is 
conformal.  We would then like to address two questions: (1) What are the equations 
of motion? (2) How do their solutions relate---if at all---to the renormalization group 
flow of the spatial theory $\bar S_{u}$ generated by the beta functions $\bar\beta^a$? 
In this paper, we answer both of these questions in the case where the tachyons are 
light, i.e. $\delta_a\equiv2-\Delta_a\ll1$.

To answer the first question we use two different methods. First we use conformal perturbation 
theory, to quadratic order in $u^a$ and lowest order in $\delta_a$.  We then do the calculation 
using a background field method, to second order in time derivatives and lowest order in 
$\bar\beta^a$. The two methods give the same equations of motion, which are written in 
equation (3.6). These equations are essentially geodesic motion for $u^a$ with respect to $\phi$, with a force term $-\bar\beta^a$. We show that for consistency $\bar\beta^a$ must be the gradient of $\bar\beta^\Phi$, which thus acts as a potential energy, with minima at the infrared fixed points. By energy conservation it would seem that the tachyon field could never settle into such a minimum, so time evolution could never give the same endpoint as RG flow. However, we show that if the dilaton is decreasing with time (which requires the spatial theory to be super-critical) the equations of motion acquire a friction term which will eventually (if our approximations do not break down) lead the couplings to settle at an infrared fixed point. The combination of the force term and the friction term thus provides the qualitative link between the time evolution and the RG flow, answering the second question.\foot{It has been known at least since \CooperVG, based on the interpretation of the time direction as the Liouville field, that in the limit of large negative linear dilaton slope the time evolution becomes identical to RG flow. The fact that, more generally, a negative linear dilaton slope leads to friction in the dynamics has been noted in several other papers on closed string tachyon condensation, including \refs{\DasDA,\SenMZ,\HeadrickHZ}.} Depending on the details of the system and the initial conditions the final state may or may not be at the same fixed point as predicted by RG flow.

In general, the final spatial theory will have a lower value of $\bar\beta^\Phi$, and therefore a lower central charge, than the initial theory. Does this not contradict the fact that the total central charge cannot change? As we will see, what happens is that the dilaton dynamically adjusts its slope so that the full theory remains critical.\foot{The same effect was calculated in a different way in a supercritical heterotic example in \HellermanQA.} In other words, central charge is exchanged between $\CC$ and $\phi$.

The plan of our paper is as follows.
In \S2, we discuss general aspects of conformal perturbation theory.
We review the computation of the one-loop beta functions, in the renormalization
scheme we adopt in this paper; we discuss
the scheme dependence of the answers; and we point out that at order $\delta_a^3$ the beta functions cannot be the gradient with respect to the Zamolodchikov metric of a scalar function.
In \S3\ we derive the beta functions
for CFTs perturbed by relevant and marginal operators,
coupled to a scalar field, using both conformal perturbation theory and the background field method.
In \S4\ we discuss solutions to these equations, both in general and in the
case of a CFT containing a unitary minimal model,
and we discuss how the dynamics appears if we change the renormalization scheme.
In \S5\ we discuss the relationship between our effective action and one computed
by tachyon scattering amplitudes; we discuss the relationship of our
theory to Liouville theory coupled to matter; and we show that in the presence
of a linear dilaton with large slope, a class of trajectories are described by the
standard renormalization group equations.  Finally, we discuss the relation of
our results to open string tachyon condensation.  In \S6\ we conclude with a speculation
regarding a Hamilton-Jacobi formulation of the tachyon dynamics.  The appendices
contain some useful technical results.  Appendix A is a review of the renormalization group
equations used to derive the beta functions.  Appendix B contains a calculation of the Zamolodchikov metric to linear order in the renormalization group scheme used in this paper.
Appendix C contains a more detailed discussion of the computation of the beta functions
by the background field method.

\newsec{Conformal perturbation theory}

In this work we are interested in perturbed two-dimensional conformal field theories 
coupled to a scalar field.  In order to develop intuition for this system,  \S2.1 is 
dedicated to a brief review of some conformal perturbation theory. In \S2.2 we discuss 
the scheme dependence of the beta functions, and in \S2.3 we address the question of 
whether they can be derived as gradients of a scalar function.

\subsec{Review}

In this subsection we review a standard calculation (\cf\ \refs{\CardyDA,\cardybook}) of 
the beta functions of a perturbed conformal field theory 
to quadratic order in the perturbations. Our starting point is an ultraviolet fixed point 
described by a nontrivial conformal field theory. This theory may or may not possess 
a Lagrangian description (the examples we have in mind are the $c < 1$ unitary 
minimal models \refs{\BelavinVU}).

If the CFT can be described by an action $S_{\rm CFT}$, we are interested in perturbed 
theories of the form
\eqn\cptaction{
	S = S_{\rm CFT} + \int d^2 z \sum_a \eps^{\Delta_a - 2} u^a \CO_a(z) = S_{\rm CFT} + S'.
}
where $u^a$ is a dimensionless coupling, $\Delta_A$ is the dimension of the spinless operator $\CO_A$ in the
 unperturbed CFT, and $\eps$ is a length scale that we will identify with the ultraviolet cutoff.  
 If a Lagrangian description of the UV theory does not exist, then we can describe the 
 perturbed theory via its correlation functions:
\eqn\corrpert{
	C_n(x_1,\ldots,x_n) = \langle \CO_1(x_1)\ldots \CO_n(x_n)\rangle_{u} \equiv
	\langle \CO_1(x_1)\ldots \CO_n(x_n) e^{ - S'}\rangle_{\rm CFT}.
}
The correlation function on the right hand side is evaluated in the CFT.

We will use the renormalization group equation:
\eqn\rgequation{
	\left( \eps\frac{\p}{\p \eps}  -\beta^c\p_c \right)
		\langle \CO_{a_1}(x_1)\ldots\CO_{a_n}(x_n)\rangle
		+ \sum_k \gamma_{a_k}^{c_k}
		\langle\CO_{a_1}(x_1)\ldots\CO_{c_k}(x_k)\ldots\CO_{a_n}(x_n)\rangle = 0.
}
Here $\beta^a$ is defined as the coefficient of $\CO_A$ in the trace of the stress tensor,
\eqn\sttrace{
\Theta = -\pi\beta^a \CO_a,
}
while $\gamma_a^b$ is the matrix of anomalous scaling dimensions, related to the beta functions by
\eqn\betatogamma{
	\p_a \beta^c = - \left(2 - \Delta_a\right) \delta^c_a - \gamma^c_a.
}
The derivation of \rgequation\ is reviewed in Appendix A.

We can compute $\beta^a$ by applying equation \rgequation\ to the partition function.  Expand the partition function out to second order in $u$:
\eqn\cptpfquad{
\eqalign{
	Z &= \left\langle e^{-\int \eps^{2-\Delta_a} u^a \CO_a}\right\rangle \cr
	 &= 
	\left\langle 1 - \int d^2z \eps^{\Delta_a- 2} u^a \CO_a(z) 
 + \half \int d^2z d^2y 
		\eps^{\Delta_b + \Delta_c - 4} u^b u^c  \CO_b(z) \CO_c(y) + \cdots\right\rangle.
}}
The calculation requires a choice of renormalization group scheme, which specifies the 
cutoff dependence. We will regulate the theory by cutting off the OPE singularities 
following the prescription of \refs{\AmitAB}:
\eqn\cptope{
	\CO_b(z) \CO_c(y) = \sum_{a} \frac{1}
		{\left(|z-y|^2 + \epsilon^2\right)^{\Delta_{abc}/2}}
		\ C^a_{bc}\ \CO_a\left(\frac{z+y}{2}\right)\ ,
}
where $\Delta_{abc} = -\Delta_a + \Delta_b + \Delta_c$.  Note that this will introduce 
cutoff dependence even when the integrated operator products in \cptpfquad\ are not 
singular.  This is a specific choice of scheme, which we dub the ``Wilsonian" scheme; 
in \S2.2 we will discuss what happens when the scheme is changed.

Setting $\bz = \half(z+y)$, $\delta z = z-y$, we can integrate over $\delta z$
to find the following:
\eqn\opeintegrate{
\eqalign{
	\int d^2 z \int d^2 y\,\CO_b(z) \CO_c(y)
& = C^a_{bc}\int_0^{|\delta z|^2=R^2 - \eps^2}\frac{d^2\delta z}{\left(|\delta z|^2 + \epsilon^2\right)^{\Delta_{abc}/2}}\int d^2\!\bz\,\CO_a(\bz) \cr
& = 4\pi C^a_{bc}\frac{\epsilon^{2-\Delta_{abc}} - R^{2-\Delta_{abc}}}{\Delta_{abc}-2}
\int d^2\bz\,\CO_a(\bz).
}}
Here the IR cutoff has been implemented in such a way that the $\eps$-dependent 
term, which gives rise to the beta function, is conveniently independent of $R$. 
Using \rgequation\ we find the following beta function:
\eqn\cptoneloopbeta{
	\beta^a = - \delta_au^a + 2\pi C^a_{bc} u^b u^c + O(u^3) \qquad
	\hbox{(no sum on $a$)},
}
where 
\eqn\deltadef{
\delta_a \equiv 2-\Delta_a
}
is the deviation of the coupling from marginality.

\subsec{Scheme dependence}

The beta function \cptoneloopbeta\ was calculated within a particular renormalization 
scheme. Changing one's scheme amounts to redefining the couplings, i.e.\ making a 
coordinate transformation on the space of couplings. More precisely, since the action 
\cptaction\ fixes the definition of the couplings at linear order, scheme changes 
correspond to coordinate transformations at quadratic and higher order:\foot{In fact there are important examples of scheme changes which are not analytic. 
This issue will not affect us in this paper, but needs to be kept in mind. We would 
like to thank T. Banks for pointing this out to us.}
\eqn\coordchange{
	u^a \to \tilde{u}^a = u^a + b_{bc}^au^bu^c + O(u^3).
}
The beta functions transform like a vector under scheme changes, so we have
\eqn\shiftedbeta{
\tilde\beta^a = 
-\delta_a\tilde u^a + 
\left(2\pi C^a_{bc}+(\delta_a-\delta_b-\delta_c)b^a_{bc}\right)\tilde u^b\tilde u^c +
O(\tilde u^3).
}
We see that the coefficient of the quadratic term is scheme dependent unless 
$\delta_a-\delta_b-\delta_c=0$. OPEs satisfying $\delta_a-\delta_b-\delta_c=0$ 
are sometimes called ``resonant", as in the discussion \refs{\ShatashviliPS}\ of boundary 
perturbations; the divergence in \opeintegrate\ is logarithmic in this case, explaining the 
scheme independence of the beta function.

One particularly simple scheme---which can only be reached if there are no 
resonant OPEs---is the one in which the beta functions are exactly linear in the 
couplings: $\beta^a = - \delta_a u^a$.  While it has the advantage of simplicity, it 
may miss interesting physics.  When the perturbations in \cptaction\ are nearly 
marginal, so that $|\delta_a| \ll 1$, the beta functions in \cptoneloopbeta\ have nontrivial 
zeros for $u^a \sim \delta_a$, which is within the realm of perturbation theory.  
These are nontrivial IR fixed points;  in string theory they are possible endpoints 
of tachyon decay.  Linearizing the beta functions pushes these fixed points off to infinite coupling.

In the remainder of this paper we will be studying string backgrounds, for which 
the full beta functions vanish.  The operators we perturb by will be nearly marginal.\foot{In principle one also has to solve the beta function equations for all the couplings that are not nearly marginal. This can be done by turning them on at order $\delta^2$, where $\delta$ is the typical deviation from marginality of the nearly marginal operators, giving a negligible ``back-reaction" on the nearly marginal couplings. In the string theory application of the next section, this corresponds to the usual integrating out of the heavy fields.} In this case, we will treat $\delta_a$ as an expansion parameter (as in \refs{\AmitAB,\LudwigRK}). If we are studying such backgrounds using conformal perturbation theory, we should perform a double expansion in $u^a$ and $\delta_a$. The scheme employed in the last subsection was such that the result \cptoneloopbeta\ was exact in $\delta_a$. Note, however, that by \shiftedbeta\ all of the terms in \cptoneloopbeta\ are scheme independent at lowest order in $\delta_a$.

Let us introduce one more interesting scheme, namely the one employed by 
Zamolodchikov in \refs{\ZamolodchikovTI}. To define it, we need to introduce 
his metric on the space of couplings
\eqn\zmetricdef{
g_{ab} \equiv \eps^{\Delta_a + \Delta_b} \langle \CO_a(\eps)\CO_b(0)\rangle
}
(note that this definition is scheme-covariant, i.e.\ $g_{ab}$ transforms as a tensor 
under scheme changes). Then his scheme, which we will denote by $\tilde u^a$, 
is defined by the condition 
\eqn\Zscheme{
\tilde g_{ab}(\tilde u) = \delta_{ab} + O(\tilde u^2).
}
In this scheme he calculates, for primary operators, the beta function to quadratic order in $\tilde u^a$,
\eqn\Zbeta{
\tilde\beta^a = -\delta_a\tilde u^a + \gamma_{bc}^a\tilde u^b\tilde u^c + O(\tilde u^3),
}
finding that the quadratic coefficient $\gamma^a_{bc}$ depends on both the OPE 
coefficient and the scaling dimensions. While he calculates this dependence exactly, 
for our purposes it is sufficient to note that the leading $\delta$ dependence occurs 
at cubic order:
\eqn\Cexpansiontwo{
\gamma_{bc}^a = 2\pi C^a_{bc}
\left(1 -\frac{1}{4}\psi''(1)(\delta_a-\delta_b-\delta_c)\left(\delta_a^2-\delta_b^2-\delta_c^2 + O(\delta^4)\right)\right),
}
where $\psi$ is the digamma function.

Comparing \Cexpansiontwo\ to \shiftedbeta, we see that Zamolodchikov's scheme is related to the Wilsonian one 
used in \S2.1 by a coordinate transformation of the form \coordchange, with
\eqn\Zcoordchagne{
b^a_{bc} = -{\pi\over2}\psi''(1)C^a_{bc}\left(\delta_a^2-\delta_b^2-\delta_c^2 + O(\delta^4)\right).
}
Using this we can transform the metric \Zscheme\ into the Wilsonian coordinate system, 
finding that the leading correction is of order $\delta^2u$:
\eqn\Wilsonmetric{
g_{ab} = \delta_{ab} + \pi\psi''(1)C^c_{ab}\delta_c^2u^c + O(u^2)
}
(where we've used the symmetry properties of the OPE coefficients of primary operators). 
In Appendix B, the metric is calculated directly to order $\delta\,u$ in the Wilsonian 
scheme, confirming the absence of a correction at that order.

\subsec{Gradient flow?}

An interesting question is whether the beta functions $\beta^a$ of a theory are in general the 
gradient of some scalar function on the space of couplings. In the paper 
\refs{\ZamolodchikovTI}, Zamolodchikov proves such a relation to second order 
in $u^a$ and lowest order in $\delta_a$:
\eqn\gradflow{
	g_{ab}\beta^b = \frac{1}{24\pi^2}\p_a C,
}
where the index on $\beta^a$ is lowered with his metric \zmetricdef, and $C$ is his $C$-function. However, this relation fails to hold at higher order in $\delta_a$. More precisely, since the cubic term in \Cexpansiontwo\ is not symmetric in the indices $a$ and $b$, the one-form $g_{ab}\beta^b$ is not closed and hence cannot be the gradient of a scalar. (This fact is implicit in the discussion in \refs{\ZamolodchikovTI}.) The problem is thus with the metric, not the $C$-function, and it would be interesting to know whether there is another metric on the space of couplings such that gradient flow does work. We should note that an exact gradient flow formula has been proven for boundary perturbations \refs{\KutasovQP,\KutasovAQ, \FriedanYC}; it has the form \gradflow\ 
with $C$ replaced by the boundary entropy $g$ and $g_{ab}$ replaced by a smeared two-point function.

\newsec{Calculation of the beta functions}

Our goal is to describe the dynamics of tachyon condensation in string theory, and the 
relation of these dynamics to two-dimensional renormalization group flows. To this end, 
inspired by the discussion in \refs{\SchmidhuberBV}, we study the following models. 
$\CC$ denotes a compact unitary CFT containing relevant spinless operators $\CO_a$ with 
dimensions $\Delta_a$. We construct a string background by tensoring $\CC$ with 
the theory of a single free boson $\phi$ representing an additional target space 
direction. There may also be a spectator CFT that contributes to the total central 
charge but otherwise will not participate in the discussion. The relevant operators 
$\CO_a$ correspond to tachyons in this background. We consider perturbations of this 
tensor product theory:
\eqn\coupledtheory{
S_{u(\phi)} = S_{\CC} + \int d^2\!z\left[-{1\over2\pi\alpha'}\p\phi\bar\p\phi + 
	\eps^{\Delta_a - 2} u^a (\phi) \CO_a\right].
}
For the sake of concreteness we have adopted a timelike kinetic term for $\phi$, 
appropriate for considering the time-dependent process of tachyon decay. All of 
our results, however, generalize straightforwardly to a spacelike direction, appropriate 
for studying spatial tachyon profiles. Because we consider only a single spacetime 
direction, the metric $G_{\phi\phi}$ can be eliminated by a gauge choice; we work 
in the gauge $G_{\phi\phi}=-1$.

We assume that the beta functions $\bar\beta^a(u)$ for the perturbed CFT absent $\phi$,
\eqn\constanttheory{
\bar S_u = S_{\CC} + \int d^2\!z\,\eps^{\Delta_a - 2} u^a \CO_a,
}
are known. Given this information, we want to know what conditions $u^a(\phi)$ must 
satisfy in order for $S_{u(\phi)}$ to be a CFT; in other words we want to know the beta 
functions $\beta^a[u^a(\phi)]$, as well as the beta function $\beta^G[u^a(\phi)]$ for 
$G_{\phi\phi}$. In this section we calculate $\beta^a$ and $\beta^G$ by two different 
methods, which have slightly different domains of applicability:
\item{(1)} In \S3.1 we use conformal perturbation theory. For this we need to assume 
that the couplings $u^a$, as well as their deviation from marginality $\delta_a\equiv2-\Delta_a$, 
are small. The results are therefore most useful in a situation in which the theory $\bar S_u$ 
has an infrared fixed point which is close to $\CC$.
\item{(2)} In \S3.2 we use a background field method. This does not require $u$ to be 
small, but does require $\bar\beta^a$ and $\phi$-derivatives of $u$ to be small.

\noindent In both cases we find the following beta functions:\foot{The reader should not 
be concerned with the fact that the RG flow generated by these beta function may not respect the gauge choice $G_{\phi\phi}=-1$; we are interested only in finding 
fixed points $\beta^a=\beta^G=0$, not in following the RG flow otherwise.}
\eqn\withoutDilaton{\eqalign{
{1\over\alpha'}\beta^a &= 
{1\over\alpha'}\bar{\beta}^a + \half\ddot{u}^a + \half\Gamma^a_{bc}\dot{u}^b\dot{u}^c, \cr
{1\over\alpha'}\beta^G &= - 4\pi^2 g_{ab}\dot u^a\dot u^b,
}}
where a dot denotes a derivative with respect to $\phi$, $g_{ab}$ is the Zamolodchikov metric of the theory $\bar S_u$,\foot{This and all other correlators in this paper are normalized. In particular, in the case of a geometrical target space the correlator contains a factor of one over the target space volume. In the infinite volume limit $g_{ab}$ therefore vanishes for localized modes. Hence what follows is valid only for bulk tachyons and tachyons that are localized in a compact space.}
\eqn\Zmetric{
g_{ab}(u) \equiv \eps^{\Delta_a + \Delta_b} \langle \CO_a(0) \CO_b(\epsilon) \rangle_u,
}
and $\Gamma^a_{bc}$ is the connection for it. 

Positivity of the Zamolodchikov metric implies that the beta functions \withoutDilaton\ 
have no fixed points aside from the static ones $u^a(\phi)=u^a_*$, where 
$\bar\beta^a(u_*)=0$. This is clearly not satisfactory, since we are interested 
in studying rolling tachyon solutions. The problem is that we have so far left 
out an important massless degree of freedom, namely the dilaton. Therefore, 
moving to a curved world-sheet, we add a Fradkin-Tseytlin term\foot{If there is a spinless operator $\CO_a$ with dimension close to zero, then one should also consider the operators $\CO_aR^{(2)}$ and $\CO_a\partial\phi\bar\partial\phi$. An example of this would be where $\CC$ is a sigma model with a large target space, in which case these operators would correspond to spatially varying dilaton and metric fluctuation. Similarly, if there is an $\CO_a$ with dimension $(1,0)$, then one should consider $\CO_a\bar\partial\phi$. We will assume that backgrounds exist such that these can be ignored to within our approximations.}
\eqn\ftterm{
S_\Phi = {1\over4\pi}\int d^2\!z\sqrt{g}\,\Phi R^{(2)}
}
to the actions $\bar S_u$ and $S_{u(\phi)}$, where in the former case $\Phi$ is a 
constant and in the latter case it is a function of $\phi$. Assuming that the dilaton 
beta function $\bar\beta^\Phi(u)$ in the theory $\bar S_u+S_\Phi$ is known, in \S3.3 
we adapt sigma model methods to compute its beta function $\beta^\Phi$ in the 
theory $S_{u(\phi)}+S_{\Phi(u)}$, as well as its contribution to $\beta^a$ and 
$\beta^G$. We find the following:
\eqn\flowequations{\eqalign{
{1\over\alpha'}\beta^a &=
{1\over\alpha'}\bar{\beta}^a + \half\ddot{u}^a + \half\Gamma^a_{bc}\dot{u}^b\dot{u}^c -\dot\Phi\dot u^a, \cr
{1\over\alpha'}\beta^G &= - 4\pi^2 g_{ab}\dot u^a\dot u^b + 2\ddot\Phi, \cr
{1\over\alpha'}\beta^\Phi &=
{1\over\alpha'}\left(\bar\beta^\Phi + {c_{\rm aux}+1\over6}\right) + \half\ddot\Phi - \dot\Phi^2.
}}
In $\beta^\Phi$ we have included a contribution $c_{\rm aux}/6$ from the ghost and 
spectator CFTs.

Like \withoutDilaton, \flowequations\ has one more equation than dynamical variable, 
so one might worry again that there will be no solutions (or only a few trivial ones). 
This is not the case for the following reason. The condition $\beta^a=\beta^G=0$ is 
sufficient for the action $S_{u(\phi)}+S_{\Phi(\phi)}$ to define a CFT. General field 
theory arguments (specifically the Wess-Zumino consistency condition) imply that 
the conformal anomaly is then a c-number rather than an operator, and hence that 
$\beta^\Phi$ is $\phi$-independent. The fact that $\beta^\Phi$ is conserved 
whenever $\beta^a=\beta^G=0$ implies that $\bar\beta^a$ and $\bar\beta^\Phi$ 
are related by
\eqn\gradientflow{
\partial_a\bar\beta^\Phi = 4\pi^2g_{ab}\bar\beta^b,
}
so that $\bar\beta^\Phi(u)$ acts as a potential function on the space of couplings $u^a$.\foot{Recall that in \S2.3 we found an obstruction to $g_{ab}\bar\beta^b$ being the gradient of a scalar function at order $\delta^3$. This is higher order in $\delta$ than we work at in the method (1) calculation, so there is no contradiction. By comparing \gradientflow\ with \gradflow\ and noting that $\bar\beta^\Phi(u^a\!=\!0)=c_{\CC}/6$, we see that, to lowest order in $\delta$, $\bar\beta^\Phi=\bar C/6$.} Equation \gradientflow\ can be proven as follows. Using $\beta^a=\beta^G=0$, we have
\eqn\gfproof{
\dot\beta^\Phi = {\partial\over\partial t}\left(\beta^\Phi - {1\over4}\beta^G\right) = \dot u^a\left(\partial_a\bar\beta^\Phi-4\pi^2g_{ab}\bar\beta^b\right).
}
For this to vanish for all $\dot u^a$ requires \gradientflow. Being a conserved 
quantity, $\beta^\Phi$ is a constraint rather than a dynamical equation of motion, and 
we have the same number of equations as variables. In section 4 we will discuss 
solutions to these equations. Note that, in view of \gradientflow, they are derivable 
from the following spacetime action:
\eqn\spacetimeaction{
{\bf S} = \int d\phi\sqrt{|G_{\phi\phi}|}e^{-2\Phi}\left(G^{\phi\phi}\left(\dot\Phi^2-\pi^2g_{ab}\dot u^a\dot u^b\right) - {1\over\alpha'}\left(\bar\beta^\Phi+{c_{\rm aux}+1\over6}\right)\right).
}

\subsec{Conformal perturbation theory method}

In this subsection the beta functions for the theory $S_{u(\phi)}$ defined in 
\coupledtheory\ are computed to quadratic order in $u$ using the following 
simple facts: (1) the conformal perturbation theory beta functions \cptoneloopbeta, \Zbeta\ are expressed in terms of the data of the CFT about which we are perturbing (dimensions of operators and OPE coefficients); and (2) if the CFT is a tensor product of two CFT's, as is the case for $S_{u(\phi)=0}$, then that data is easily expressed in terms of the data of the two factors. The result will be the beta functions given in \withoutDilaton.

It will be convenient for us to work in Zamolodchikov's scheme, introduced in \S2.2, 
where according to \Zscheme\ the connection $\Gamma^a_{bc}$ on the space of 
couplings is of order $u$. Once we establish that \withoutDilaton\ holds in this 
``normal" coordinate system, general covariance on the space of couplings 
demands that it hold in any coordinate system. 

The unperturbed CFT is a tensor product of $\CC$ and the Gaussian model 
for $\phi$. The scaling operators for the product theory are products of scaling 
operators in each factor. Hence we expand the operator $u^a(\phi)$ in scaling 
operators $e^{ik\phi}$ (as always the $\CO_a$ are scaling operators), writing:
\eqn\couplings{
u^a(\phi)\CO_a = \int dk\,\tilde u^a(k)\CO_{ka}, \qquad \CO_{ka} \equiv e^{ik\phi}\CO_a,
}
and the index $a$ used to label the scaling operators in Section 2 is replaced 
by a double index $ka$. In a product theory, scaling dimensions add and OPE 
coefficients multiply, so we have
\eqn\factorization{
\delta_{ka} = -\half\alpha'k^2 + \delta_a, \qquad
C^{k_1a}_{k_2b,k_3c} = \delta(k_1-k_2-k_3)C^a_{bc}.
}
Using \Zbeta, we therefore have
\eqn\momentum{
	\tilde\beta^a(k) = 
-\left(\half\alpha'k^2+\delta_a\right)\tilde u^a(k) + 2\pi C^a_{bc}\int dk'\,\tilde u^b(k')\tilde u^c(k-k'),
}
or in position space
\eqn\position{
\eqalign{
	\beta^a & = 
	\half\alpha'\ddot u^a - \delta_au^a + 2\pi C^a_{bc}u^b u^c \cr
	& = \half \alpha' \ddot u^a + \bar \beta^a.
}}
Note that there are order $\delta^3$ corrections to both equalities of \position.

The Fourier transform $\tilde G_{\phi\phi}(k)$ of the spacetime metric is the coupling for the 
operator
\eqn\khatdef{
\CO_{\hat k} \equiv {1\over2\pi\alpha'}\p\phi\bar\p\phi e^{ik\phi}.
}
This operator has dimension $2-\half\alpha'k^2$; however, since we are working in a gauge 
where $G_{\phi\phi}$ is constant the first term on the right-hand side of \cptoneloopbeta\ 
vanishes. In this gauge the leading contribution to the beta function will be from the 
quadratic term, for which we need the OPE coefficient $C^{\hat k_1\bf 1}_{k_2a,k_3b}$, 
where $\bf 1$ denotes the unit operator in $\CC$. This again factorizes into a product of 
OPE coefficients in the respective theories, which are straightforwardly computed:
\eqn\GOPE{\eqalign{
C^{\hat k_1\bf 1}_{k_2a,k_3b} &= C^{\hat k_1}_{k_2k_3}C^{\bf 1}_{ab}, \cr
C^{\hat k_1}_{k_2k_3} &= -{\pi\over2}\alpha'(k_2-k_3)^2\delta(k_1-k_2-k_3), \cr
C^{\bf 1}_{ab} &= g_{ab}.
}}
Thus we have
\eqn\betaGfirst{
\beta^G = 
\pi^2\alpha'g_{ab}\left(\ddot u^au^b+u^a\ddot u^b-2\dot u^a\dot u^b\right).
}
The terms in \betaGfirst\ with second $\phi$ derivatives can be removed by a 
diffeomorphism\foot{The reason for this diffeomorphism is not entirely clear to us, 
but it is necessary for agreement with the calculation of \S3.2, as well as the tachyon 
beta functions calculated in \refs{\FradkinYS,\FradkinPQ,\CallanIA,\DasCZ,\JainJV,\CooperVG}.} 
with $\delta\phi\sim g_{ab}\p_\phi(u^au^b)$, leaving
\eqn\betaGsecond{
\beta^G = 
-4\pi^2\alpha'g_{ab}\dot u^a\dot u^b.
}
The diffeomorphism will add to $\beta^a$ a term $\dot u^a\delta\phi$, which is cubic in $u$ and therefore higher order than what we have calculated.

\subsec{Background field method}

In this subsection we will derive the beta functions \withoutDilaton\ using a background 
field expansion in $\phi$, as in \refs{\AlvarezGaumeHN,\CallanIA}. We write 
$\phi(z) = \bar{\phi} + \xi(z)$, where $\bar{\phi}$ is constant and $\xi$ is a small 
fluctuation. We then Taylor expand the trajectory, 
$u^a(\phi) = \bar u^a + \dot{u}^a\xi + \half\ddot u^a\xi^2 + \cdots$, where 
$\bar u^a=u^a(\bar\phi)$, $\dot u^a=\dot u^a(\bar\phi)$, etc. In the background 
field method we treat $\dot u^a$ and $\ddot u^a$ as small coupling constants, 
whereas $\bar u^a$ may be an arbitrary point in the space of couplings. We decompose 
the action \coupledtheory\ into the action for the static theory $u(\phi)=\bar u$, plus 
perturbations parametrized by $\dot u^a$ and $\ddot u^a$:
\eqn\actionexpand{
	S_{u(\phi)} = S_{\bar u} + \int d^2z\,\eps^{\Delta_a-2}\left[\dot{u}^a\xi + \half\ddot u^a\xi^2 + \cdots\right]\CO_a.
}

The derivative expansion will work so long as the beta functions are small all along the flow, 
that is, $\bar\beta^a(u) \sim \delta \ll 1$. (Note that this $\delta$ is not the same as the 
deviation from marginality $\delta_a$ used in the previous subsection; however,
it will in general be of the same order.) Our equations will imply 
that $\ddot u$ is also of order $\delta$. For clarity, we will first do the calculation in the 
case that the $u^a$ parametrize a moduli space of CFTs, that is $\bar\beta^a(u)=0$. We will 
then add in the effects of non-zero beta functions.

The expansion of the partition function to second order in $\xi$ is:
\eqn\pertexpansion{
\eqalign{
	Z & = \left\langle 1 - \int d^2z\,\dot u^a\xi(z)\CO_a(z)
	- \half \int d^2 z\,\ddot{u}^a\xi^2(z) \CO_a(z) \right. \cr
	& \qquad\qquad\qquad\qquad {} + \left. \half \int d^2 z\int d^2 w\,\dot{u}^a\dot{u}^b\xi(z)\xi(w)
		\CO_a(z)\CO_b(w)
	 + \cdots\right\rangle_{\epsilon,\bar u},
}}
where expectation values are evaluated in the static theory \coupledtheory\ 
with $u^a(\phi) = \bar u^a$. The OPEs of exactly marginal operators are 
constrained to have the following form \KutasovXB:
\eqn\marginalopes{
\CO_a(z)\CO_b(w) = {g_{ab}\over|z-w|^4} + \Gamma_{bc}^a\CO_a(z)\delta^{2}(z-w) + \cdots
}
(in particular, for exactly marginal operators $C^a_{bc}$ must vanish, otherwise according 
to \cptoneloopbeta\ $\bar\beta^a$ would be non-zero at second order in $u-\bar u$). 
Here $g_{ab}$ is the Zamolodchikov metric and $\Gamma^a_{bc}$ its connection 
at the point $\bar u^a$ in coupling constant space. Inserting \marginalopes\ into 
\pertexpansion, we find
\eqn\Zexpand{
\eqalign{
	Z & = \left\langle 1 - \int d^2z\,\dot u^a\xi(z)\CO_a(z)
	- \half\left(\ddot u^a+\Gamma^a_{bc}\dot u^b\dot u^c\right)\int d^2 z\,\xi^2(z) \CO_a(z) \right. \cr
	& \qquad\qquad\qquad\qquad\qquad\qquad {}+ \left.\half g_{ab}\dot u^a\dot u^b\int d^2z\int d^2w {\xi(z)\xi(w)\over|z-w|^4}
		 + \cdots\right\rangle_{\epsilon,\bar u}.
}}
The beta functions can be extracted from the logarithmically divergent parts of this 
expression. We can regulate $\xi^2(z)$ by point-splitting, giving
\eqn\pointsplit{
\xi(z)\xi(z+\eps) = {}:\xi(z)\xi(z+\eps): + \alpha'\ln\eps,
}
yielding the beta function
\eqn\betaa{
\beta^a = {\alpha'\over2}\left(\ddot u^a+\Gamma^a_{bc}\dot u^b\dot u^c\right).
}
The logarithmically divergent part of the last term in \Zexpand\ can be isolated by writing 
$|z-w|^{-4}=-\partial_z\partial_{\bar w}|z-w|^{-2}$ and integrating by parts with respect to 
$z$ and $\bar w$. We obtain
\eqn\forbetaG{
\half g_{ab}\dot u^a\dot u^b\int d^2z\int d^2w {\partial\xi(z)\bar\partial\xi(w)\over|z-w|^2}.
}
When regulated, either with a sharp cutoff or in the manner of \S2.1, this divergent 
integral leads to the beta function
\eqn\betaG{
\beta^G = -4\pi\alpha'g_{ab}\dot u^a\dot u^b.
}

Once the dilaton is included (as it will be in the next subsection), the beta functions 
$\beta^a$ and $\beta^G$ can be derived from the spacetime action \spacetimeaction\ 
with $\bar\beta^\Phi=c_\CC/6$. Hence our calculation agrees with the well-known fact 
that for moduli the coefficient of the kinetic term in the spacetime action is given by the 
Zamolodchikov metric.

We now generalize to the case where the $u^a$ are nearly but not exactly moduli, by 
considering 
corrections to \betaa\ and \betaG\ due to small $\bar\beta^a$. The full beta function can be 
calculated in an expansion in derivatives and in $\delta$. A more careful discussion
of the beta functions, and of this expansion, appears in Appendix C. 
Here we simply note that the full beta functions to zeroth order
in $\delta$ and to second order in derivatives will be \betaa\ and \betaG\ as given above.
As we can see from \actionexpand, 
the calculation to zeroth order in derivatives and to first order in $\delta$
is computed using the action $S_{\bar u}$ and just gives us $\bar\beta$.  The sum of these
two contributions is given in \withoutDilaton.  

In the end we are interested in the case $\beta^a = \beta^G = \beta^{\Phi} = 0$.
Typical solutions will tie together the derivative expansion and the expansion in
$\delta$: that is, we can take $\ddot u$ and $\dot u^2$ to be of order $\delta$.

\subsec{Inclusion of the dilaton}

The final step in our calculation is to include the effects of the dilaton.  As we will see in \S4,
the dilaton is unavoidably excited during tachyon condensation.

We will be considering dilaton couplings of the form \ftterm.  This means that we are ignoring
more general dilaton profiles of the form 
\eqn\gendil{
	S'_{\Phi^a} = \int d^2 z \Phi^a(\phi) \CO_a R^{(2)}.
}
This makes sense so long as the perturbations in \coupledtheory\ correspond to nearly marginal
operators whose OPEs close only on other nearly marginal operators.  In this case 
couplings of the
form \gendil\ will be highly irrelevant, and can be ignored.

The lowest-order contribution to the dilaton beta function comes from the conformal
anomaly of the UV fixed point (including the contribution of $\phi$)
and the dimension of $\Phi(\phi)$:
\eqn\dilbetaone{
	\beta^{\Phi} = \frac{c_{\rm aux} +1}{6} + \half \alpha' \ddot \Phi.
}
In principle, computing the effects of the dilaton on $\beta^a$ and $\beta^G$, and the 
higher order contributions to $\beta^{\Phi}$, requires working on a curved world-sheet. 
Instead, we follow the discussion in \refs{\TseytlinTT,\TseytlinWS,\CallanJB}, and note 
that these terms arise from the improvement term in the stress tensor due to 
$S_{\Phi(\phi)}$. In particular, the trace of the stress tensor receives the following extra term:
\eqn\improvement{\eqalign{
\Theta_{\rm improve} &= -{1\over\alpha'}\partial\bar\partial\Phi(\phi) \cr
&= -{1\over\alpha'}\left(\dot\Phi\partial\bar\partial\phi+\ddot\Phi\partial\phi\bar\partial\phi\right) \cr
&= \pi \dot\Phi\dot u^a\CO_a + {1\over4\alpha'}\dot\Phi^2R^{(2)}-{1\over\alpha'}\ddot\Phi\partial\phi\bar\partial\phi,
}}
where in the last line we've used the world-sheet equation of motion for $\phi$, which 
is of course satisfied by the operator $\Theta$. The corresponding terms in the beta 
functions can be read off using \sttrace: $\beta^a_{\rm improve}=-\dot\Phi\dot u^a$, 
$\beta^G_{\rm improve}=2\ddot\Phi$, $\beta^\Phi_{\rm improve} = -\dot\Phi^2$.

\newsec{Solutions to tachyon equations of motion}

In the previous section we derived the equations of motion \flowequations\ for the tachyons. 
In this section we would like to study solutions to these equations. In \S4.1\ we will describe 
various trajectories corresponding to the time evolution of a tachyon condensate, driven by 
various types of potential energy functions. In \S4.2\ we discuss the case in which $\phi$ is 
a spacelike direction and the $\phi$-dependent tachyon profile describes a domain wall.

An explicit example which can be studied using our methods
is a product CFT with one factor described by a
$c < 1$ unitary minimal modes \refs{\BelavinVU,\FriedanXQ}.\foot{The equations of
motion for this system, and some qualitative features of their solution, 
were also described by Sen \refs{\SenMZ}.} We will review this system
and the corresponding tachyon dynamics in \S4.3.  

In \S4.4\ we will discuss the issue of scheme dependence for these trajectories.

\subsec{Tachyon evolution for a variety of potentials}

Let us collect the beta functions derived in Section 3:
\eqn\fullbeta{\eqalign{
{2\over\alpha'}\beta^a &=
\ddot{u}^a + \Gamma^a_{bc}\dot{u}^b\dot{u}^c -2\dot\Phi\dot u^a + g^{ab}\partial_bV, \cr
{1\over2\alpha'}\beta^G &= - g_{ab}\dot u^a\dot u^b + \ddot\Phi, \cr
{1\over\alpha'}\beta^\Phi &= V + \half\ddot\Phi - \dot\Phi^2,
}}
where for convenience we define
\eqn\Vdef{
V(u) \equiv {1\over\alpha'}\left(\bar\beta^\Phi(u) + {c_{\rm aux}+1\over6}\right),
}
and redefine $g_{ab}$ by a factor of $2\pi^2$: $g_{ab}=2\pi^2\eps^{\Delta_a + \Delta_b} \langle \CO_a(0) \CO_b(\epsilon) \rangle$.

As discussed at the beginning of Section 3, $\beta^\Phi$ is conserved when 
$\beta^a=\beta^G=0$. In order to eliminate the second derivative term in 
$\beta^\Phi$, we define
\eqn\hamiltonian{
H \equiv {1\over\alpha'}\beta^\Phi-{1\over4\alpha'}\beta^G = V - \dot\Phi^2+{1\over2}g_{ab}\dot u^a\dot u^b.
}
$\beta^a = 0$ and $\beta^G = 0$ can then be considered as the equations of motion for the 
tachyon and dilaton respectively, and $H=0$ as the Hamiltonian constraint.


The question we would like to address in this subsection is the following: 
Imagine we are given the positions and velocities of $u^a$ and $\Phi$ at some 
initial time $\phi_{\rm i}$ (possibly $\phi_{\rm i}=-\infty$), satisfying $H=0$ and 
any other constraints. What can we predict for the long-time behavior of the system?\foot{There is some overlap between this subsection and the paper \TseytlinCD\ by Tseytlin. In particular, the solution (4.4) for motion on a moduli space first appeared there. (We would like to thank the referee for pointing this reference out to us.) Also, the paper \SuyamaWD\ by Suyama, which appeared shortly after the first version of this paper, has substantial overlap with this subsection.}

The first thing to note is that it is not possible to decouple the dilaton when the 
tachyon is evolving. In particular, given the positive definiteness of the Zamolodchikov 
metric, $\beta^G = 0$ implies that $\ddot\Phi>0$ as long as $\dot u^a \neq 0$.  
The second point is that the qualitative features of the dynamics depend strongly 
on the initial value of $\dot\Phi$.  The term $-2\dot\Phi\dot u^a$ in $\beta^a$ 
implies that $\dot\Phi < 0$ leads to damped motion for $u$, while $\dot\Phi > 0$ leads
to anti-damped motion. Since $\ddot\Phi>0$, once the motion is anti-damped it 
will always be 
anti-damped. Anti-damped motion is unpredictable, and will quickly leave the 
regime of validity of \fullbeta. Therefore we must restrict ourselves to trajectories 
for which $\dot\Phi \leq 0$ at all times.

Let us consider the case that $V$ is constant, i.e.\ $\bar\beta^a(u)=0$. In this case 
$u^a$ parametrizes  a moduli space of CFTs, which are supercritical, critical, or 
subcritical depending on the sign of $V$. According to the equation $\beta^a = 0$, 
$u^a(\phi)$ will follow a geodesic of $g_{ab}$, with a speed that decreases due to 
the dilaton friction.  If we parametrize the geodesic by the proper distance $u$ from 
the initial point, then the trajectory $u(\phi)$, $\Phi(\phi)$ can be computed exactly. 
In the super-critical case $V>0$, the solution is
\eqn\vpositive{\eqalign{
u(\phi) &= u_0 - \sqrt{2}\,{\rm arctanh}\exp\left(-2\sqrt V(\phi-\phi_0)\right), \cr
\Phi(\phi) &= \Phi_0 - {1\over2}\ln\sinh\left(2\sqrt V(\phi-\phi_0)\right),
}}
where the parameters $\phi_0$, $u_0$, and $\Phi_0$ are fixed by the initial conditions. 
We see that $u$ travels a finite distance as $\phi\to\infty$. On the other hand, in the 
critical case $V=0$, $u$ slows down but travels an infinite distance as $\dot\Phi$ 
asymptotically approaches zero from below:
\eqn\vzero{
u(\phi) = u_0 + {1\over\sqrt2}\ln(\phi-\phi_0), \qquad
\Phi(\phi) = \Phi_0 - {1\over2}\ln(\phi-\phi_0).
}
In the sub-critical case $V<0$, the Hamiltonian constraint shows that $\dot u^2$ can 
never go to zero; instead, as it decreases $\dot\Phi$ passes through zero and becomes 
positive, after which the motion becomes anti-damped.

Now let us consider the more interesting case where $V(u)$ is not constant. As above, 
$u^a$ cannot come to a stop in a region with $V < 0$. If it remains in a region for which 
$V \geq 0$, then it will eventually settle into a minimum of $V$. In particular, we can 
have a transition from a local maximum of the potential at $\phi=-\infty$ to a local 
minimum at $\phi=+\infty$. In this time-dependent tachyon condensation process, 
leading from a super-critical CFT to a less super-critical (or critical) CFT, the dilaton 
responds so that in the final state it is linear with the correct slope to make the full 
theory critical. For this class of damped trajectories, the story is similar to standard
RG flows, in that the trajectory interpolates between RG fixed points. There will be 
qualitative differences at intermediate times; in particular, the tachyon may (depending 
on the height of the potential and the strength of the damping) oscillate about the 
minimum before settling down. Furthermore, whether the endpoint is the same as the IR fixed point of the RG flow depends on the details of the potential landscape and the strength of the damping. However, in the limit of infinite damping (infinitely large linear dilaton slope), the trajectory is clearly identical to the RG flow, as can be argued by considering the time direction as a Liouville field \CooperVG.

We should note that when studying tachyon evolution using conformal 
perturbation theory, this approximation is quite fragile.  If the tachyon mixes 
with marginal or nearly marginal operators such as additional graviton or dilaton modes 
(which will appear in our formalism as 
operators of the form $\CO_a$ and $\CO_a R^{(2)}$), these modes 
can become large. This can be seen, for example, in \refs{\DineCA}. 
On the other hand, as long as $V$ and $\p_b V$ remain small, the derivative 
expansion is still a good one.

As we have emphasized, the equations of motion \fullbeta\ are valid when the beta functions (i.e.\ gradients of $V$) are small. An example of a situation where this occurs, namely along the RG flows connecting the minimal models, will be discussed in subsection 4.3 below. But what happens in the more common situation where the tachyon masses (and hence beta functions) are of order one in string units? In one direction, if we take the spatial theory to be highly supercritical and give the dilaton a large negative slope, then we still have the result of \CooperVG\ that the time evolution will follow the RG flow (their argument does not depend on having small beta functions). In the other direction, if we start from a critical theory (or a supercritical one with a positive dilaton slope) the anti-damping observed above suggests (but of course does not prove) that the dynamics will lead either to a singularity or to strong string coupling in finite time.


Finally, we remind the reader that in this paper we deal exclusively with bulk tachyons or tachyons that are localized in a compact space. Let us make a few comments regarding tachyons localized in non-compact spaces. First, as mentioned in footnote 9 above, as the target space volume goes to infinity, the Zamolodchikov metric components $g_{ab}$ for localized operators go to zero, so the dilaton equation of motion reduces to $\ddot\Phi=0$. (Keep in mind that here $\Phi$ represents the zero mode of the dilaton; localized modes may well get excited and play an important role in the dynamics \AdamsSV.) As above, $\dot\Phi$ will be zero or non-zero depending on whether the spatial theory is critical or supercritical, but in this case there will be no back-reaction on it (essentially, it is locked down by the boundary conditions at infinity). $\dot\Phi$ will still serve to damp or anti-damp the tachyon dynamics (via the term $-2\dot\Phi\dot u^a$ in $\beta^a$); in particular for large negative values of $\dot\Phi$ the time evolution will mimic RG flow. Finally let us note that, unlike in the bulk tachyon case, here there is no obstruction, either in RG flow or in time evolution, for an initially critical background to go to another critical one: no obstruction for RG flow because the $C$-theorem doesn't apply to non-compact target spaces, and no obstruction for time evolution because $\dot\Phi$ will remain zero at all times.

\subsec{Domain walls}

If $\phi$ is a spacelike variable then the flow of $u^a, \Phi$ in $\phi$ will describe a 
domain wall solution.  One example of such a background is a perturbed 
conformal field theory with subcritical central charge, coupled to a Liouville 
field. For such so-called ``Liouville flows" \refs{\DavidVM,\HeadrickHZ}, all the terms in \fullbeta\ and \hamiltonian\ change sign, except those involving $V$. The discussion above therefore 
holds with the same words if we simply change the sign of $V$. A specific example is the ``hairy black hole" constructed in \GiddingsDE.

\subsec{Review: Minimal models coupled to a scalar field}

We would like to give the reader an explicit example of a set of
perturbed CFTs that are within reach of our approximations.
Recall \refs{\BelavinVU,\FriedanXQ}\ that there are a set of bosonic $c < 1$ 
solvable unitary "minimal" CFTs $M_p$ labelled by integers $p = 3,4,5,\ldots$ with central
charge 
\eqn\minimalc{
	c_p = 1 - \frac{6}{p(p+1)}\ .
}
These have a Landau-Ginzburg description in two dimensions \refs{\ZamolodchikovDB,\KastorEF},
as the IR limit of a 2d scalar field theory with a polynomial potential:
\eqn\lgaction{
	S = \int d^2z \left[ \p\varphi\bar \p \varphi + g \varphi^{2(p-1)}\right]\ .
}

A subset of the conformal primaries of $M_p$ have a description as powers of $\varphi$ in
the Landau-Ginzburg description.  One such operator is 
the least relevant field $\CO = \varphi^{2p - 4}$ with dimension
$$\Delta = 2 - \frac{4}{p+1}$$
Let $u$ be the dimensionless coupling to this field.  Then in the Wilsonian scheme, 
\eqn\betafunction{
	\bar\beta_u = - \left( \frac{4}{p} + O(\frac{1}{p^2})\right) u + \pi \left(\frac{4}{\sqrt{3}} - O(\frac{1}{p})\right) u^2 \equiv \eps u + \pi b u^2
}
A perturbation by $\CO$ flows under the renormalization group to a new IR fixed point at
$u_{IR} = \frac{\eps}{\pi b}$, which is well within the realm of perturbation theory for
$\eps \ll 1, p \gg 1$.  We might expect from the Landau-Ginzburg description of this
operator that this RG flow corresponds to the flow $p \to p - 1$.  This has been checked
by computing the central charge at the new fixed point \refs{\ZamolodchikovTI,\LudwigGS}.
Note that if we expand $\beta$ about the new fixed point, $u = u_{IR} + \delta$,
the beta function is
\eqn\irbeta{
	\bar\beta_{\delta} = \eps \delta + \pi b u^3
}
The leading term indicates that about this new fixed point, the corresponding operator is
irrelevant, as Zamolodchikov has shown \refs{\ZamolodchikovTI}. In this paragraph
we have not added the additional constant term equal to the central charge of $M_p$.

Now couple this theory to a scalar field and let $u = u(\phi)$.  
The equations of motion are:
\eqn\lindilminimal{
	0 = \ddot u - \dot \Phi \dot u - \eps u + \pi b u^2 + \cdots.
}
This describes the motion of a particle in the potential
\eqn\minimalpotential{
	V(u) = -\half \eps u^2 + \frac{\pi b}{3} u^3 + {\rm constant}.
}
The full potential will include a constant term from the central charge of
the other CFT factors in the string background. If
$V > 0$ when $u=u_{IR}$, and $\dot\Phi < 0$, then the tachyon will
interpolate between the RG fixed points $u = 0$ and $u = u_{IR}$,
perhaps oscillating about $u = u_{IR}$ before settling there at $\phi\to\infty$.
As described in \S4.1, the dilaton slope will adjust itself
in this limit to maintain a critical central charge.\foot{This
background has been constructed in closed string field theory
in \refs{\MukherjiTB}.}

In this case, we expect that both conformal perturbation theory and the derivative expansion remain valid for the class of damped trajectories discussed above, since 
in this case the depth of the potential is of order $\delta^3$ and its derivatives are order $\delta^2$.

\subsec{Scheme dependence}

Other schemes may appear simpler, but
obscure interesting physics.  For example, absent ``resonant" 
OPEs, we may perform a change of coordinates $u\to \tilde{u}(u)$
which linearizes $\bar\beta$.   In examples such as that in \S4.2, this will push the IR fixed
point $u_{IR}$ off to infinite coupling.  Nonetheless, our analysis above indicates for a class
of damped trajectories, the system should reach $u_{IR}$ in finite time and oscillate 
about this point.  This will be reflected by the fact that the transformation of the $\ddot u$
term will induce large (of order $1/\delta$) terms proportional to $\dot{\tilde u}^2$.  
The velocity of $\tilde u$ will have to become infinite.  

We may also try to linearize $\beta^a$ by choosing $U(u,\Phi)$ such that
\eqn\linearized{
	\beta^U = - \ddot U + \delta\,U
}
If $U$ begins near $0$ with positive velocity, it will evolve monotonically to infinity
in infinite time, completely obscuring the
behavior described in \S4.1.  This coordinate transformation must be highly 
nonlocal in time.

\newsec{Relation to known results}

In this section we will explore the relationship between our 
calculations and other pictures of tachyon dynamics.
In \S5.1\ we will compare our spacetime effective potential computed in
conformal perturbation theory to the result computed from
tachyon scattering amplitudes in the case of nearly massless tachyons.
In \S5.2\ we will discuss this relationship between our solutions, 
backgrounds of Liouville theory coupled to matter,
and standard renormalization group flows.  In \S5.3\ we will
review an argument \refs{\SchmidhuberBV,\CooperVG}\ that for
linear dilaton backgrounds with large slope, there is a class of trajectories that
are described by the standard renormalization group equation $\dot u = \bar\beta$.
Finally, in \S5.4\ we will discuss the well-studied case of open string tachyon decay,
and how this would appear in our picture.

\subsec{Effective action from scattering amplitudes}

In conformal perturbation theory, our equations of motion can be derived
from an effective action whose potential is the Zamolodchikov $C$-function.
(In the derivative expansion of \S3.2, this is merely a plausible conjecture).
This result is hardly surprising, and for nearly marginal $\CO_a$
has been noted, for example, in section 15.8 of \refs{\PolchinskiRR}.
For such light tachyon fields, the tree-level effective action can be
probed by computing scattering amplitudes, isolating the potential by
expanding the amplitudes in $\delta_a$ and in $\phi$-derivatives.
The mass term in such cases is automatic.  The cubic term
$\delta V \sim C_{abc}u^a u^b u^c$ is guaranteed up to an overall
coefficient, as the three-point function will be proportional to the OPE
coefficients of the theory.

\subsec{Relation to 2d gravity in conformal gauge}

When the dilaton is linear in $\phi$,
$\Phi = -\half Q\phi$, the sum of \coupledtheory\ and \ftterm\ is the ansatz 
\refs{\DavidHJ,\KnizhnikAK}\
for 2d gravity (i.e.\ Liouville theory) in conformal gauge,
coupled to a CFT with central charge $c = 25 - 3 Q^2$.  
This fact is partial motivation for the statement that there should
be a relationship between the time evolution equations \fullbeta\ and
the renormalization group flow of the theory \cptaction. 
We would like to discuss the relation of our solutions to existing lore about Liouville theory.

Recall that in a derivative expansion, the beta function equations are:
\eqn\linearbeta{
	\beta^a =  \ddot u^a +Q \dot{u}^a + \bar \beta^a = 0.
}
(We set $\alpha'=2$, and neglect the connection term.) Let us consider the case where 
$u$ is small, so that $\bar\beta^a \sim -\delta_a u^a$.
Eq. \linearbeta\ has the solution \refs{\DavidHJ,\KnizhnikAK}
\eqn\linbetasolution{
	u^a(\phi) = d^a_+ \exp\left(\alpha_+^a \phi\right) 
		+ d^a_- \exp \left(\alpha_-^a \phi\right)
}
where
\eqn\branches{
	\alpha_{\pm}^a = \frac{Q}{2} \pm \sqrt{\frac{Q^2}{4} + \delta_a}\ .
}
These solutions define the ``gravitational dressing" of $\CO_a$.\foot{These formulae were used for the case $Q=0$ to describe the onset of tachyon condensation in the critical dimension in \refs{\SuyamaHW}.} Note that because $\phi$ is timelike, the sign in front of $\delta_a$ is opposite that of the standard
gravitational dressing formula.

The existence of these two branches of the gravitational dressing formula
are the result of \linearbeta\ being second order in time derivatives. From the
standpoint of time-dependent backgrounds, both should be kept.
In studies of Liouville theory, it was argued that the solutions $u\sim e^{\alpha_+ \phi}$
should be discarded from the theory \refs{\SeibergEB,\PolchinskiMH,\GinspargIS}.  
The requirement that operators
$e^{\alpha\phi}$ appearing in the action must satisfy $\alpha \leq \frac{Q}{2}$
is known as the ``Seiberg bound".  This bound was based on several criteria.  First,
in Liouville theory, $\phi$ is the conformal factor, and operators $e^{\alpha_- \phi} \CO$
grow or decay in the infrared when $\CO$ is a relevant or irrelevant operator, respectively.
Secondly, typical studies of Liouville theory focus on $c < 1$ matter coupled to Liouville
gravity.  The Liouville field is a spacelike direction and the cosmological constant operator
$e^{\gamma\phi}$ acts as a Liouville wall", truncating the spectrum to operators
satisfying the Seiberg bound. Finally, calculations in the matrix model for 2d matter
coupled to gravity matched Liouville theory calculations with the Liouville
dependence of $u^a$ satisfying this bound.

Not all of these arguments have force.  For the second point,
when $\phi$ is a timelike direction,
one typically does not truncate the spectrum. One merely sets initial conditions
at some fixed $\phi_0$ and evolves the background forward. The two branches
reflect the fact that the low-energy target space dynamics is governed by 
second-order equations.  We will discuss an open string example below
where the two branches are crucial to describe the complete physics
of tachyon condensation.  For the final point, sources for ``wrong branch" modes have
been identified as being sourced by a specific type of 
D-brane in 2d string theory backgrounds, and given a
matrix model interpretation \refs{\KutasovFG}.  The first point, regarding the
renormalization group prescription, is more confusing.  Typically one
does not specify the initial ``velocity" $\Lambda\p_{\Lambda} u$ in
RG flows.  We will return to this point in the conclusion, and offer a suggestion.

\subsec{Recovering first-order RG flows}

If one begins with a nonconformal theory, or a conformal theory with $c \neq 26$, 
and couples it to 2d gravity, the theory should become
conformal when the scale factor of the metric is integrated over.
This scale factor acquires dynamics and is described in conformal gauge
by the Liouville field.  The coupling of operators to this field should
be related to the underlying RG flow of \cptaction, as the latter is
precisely a function of the scale dependence of the theory.

The differences between tachyon dynamics and standard RG trajectories are clear from
the previous section.  First, if $e^{2\phi}$ is the scale factor of the metric,
one would expect couplings with dimension $-\delta$ to depend on $\phi$ as
$\delta S \sim \int d^2x e^{\delta\phi} u^a\CO_a$.  The actual formulae \linbetasolution,\branches\
are more complicated even when $Q\sim 0$.  More strikingly, the RG equations are
first order in time.\foot{One response people have offered to us is to draw
an analogy to the Hamilton-Jacobi formulation of the ``holographic renormalization
group" \refs{\BalasubramanianJD,\deBoerXF}\ in the context of the AdS/CFT correspondence.
In this formulation the evolution equations naively appear to be first order
in time.  We will address this analogy in our conclusion.}

Nonetheless, we can see in \S4.1\ that for a class of trajectories, there are qualitative
similarities.  For systems such as those described in \S4.2, anti-damped motion
leads to a relaxation to an IR fixed point as $\phi\to \infty$. 

More generally, we recover motion that is first order in time when
the equations \fullbeta\ satisfy a "slow roll" condition such that some term
in the $\Gamma \dot u^2 - \dot \Phi \dot u$ part of the
tachyon beta function dominates the other two-derivative terms.  
This can occur when one coupling,
such as the dilaton, has a first derivative larger than the others, but 
small higher derivatives. A related example is inflaton dynamics in
standard slow roll inflation; there the coupling of the scale factor
to the inflaton dominates the inflaton equations of motion,
and so long as this scale factor itself changes slowly, the classical
inflaton dynamics are effectively first order in time.\foot{A.L. would
like to thank Gary Shiu for reminding him of this analogy.}

As a particular example, imagine a 2d CFT coupled to 2d gravity, very close to a fixed
fixed point of the CFT.  Let the
Liouville field at zeroth order be described by a linear dilaton profile $\Phi = Q\phi$,
with $Q$ small enough for the derivative expansion to be valid,
and perturb the CFT by a nearly relevant operator so that $\delta \ll Q$.  We can self-consistently
choose a set of trajectories such that the terms $\ddot u + \Gamma\dot u^2$ terms
in $\beta^a$ are negligible.  The approximate evolution equation is:
\eqn\gdfirstorder{
	Q \dot u = -\bar\beta \sim \delta u + \ldots 
}
The second derivative terms are of order $\delta^2$.  For small $u$, when $\bar\beta$
can be treated as linear, this corresponds to choosing the ``allowed" branch $\alpha_-$ of
the gravitational dressing formula. This equation
leads to the "gravitationally dressed scaling dimensions"
found by \refs{\KlebanovMS}\ in light cone gauge.  This was pointed out
by \refs{\SchmidhuberBV}\ and the argument runs as follows.
The natural physical scale in such models is the 2d cosmological constant.
This is the gravitationally dressed coupling to the identity operator in the CFT:\foot{This
operator is problematic in our approximation scheme---if it appears in the 2d action, 
it will only have small values and small $\phi$ derivatives for large negative $\phi$.  Even 
this requires fine tuning
so that the ``wrong branch" dressing does not also appear.}
\eqn\gdcosmo{
	\Lambda^2 = e^{\gamma\phi}
}
For this to have dimension 2, we can choose the value
\eqn\ccdressing{
	\gamma = {Q\over 2} - \sqrt{{Q^2\over 4} + 2}
}
corresponding to the allowed branch.  We can then define
\eqn\newderiv{
	\frac{\p}{\p\phi} = \frac{\gamma}{2}\Lambda\frac{\p}{\p\Lambda}
}
At lowest order in $u$, the effect is to change the effective scaling dimension
of $\CO$ to $\tilde\delta$, defined via:
\eqn\firstordertwo{
	\Lambda\p_{\Lambda} u = \frac{2}{\gamma Q} \bar\beta \sim
	- {2 \delta \over \gamma Q}  u \equiv -\tilde \delta u
}
This matches the light cone gauge calculation in \refs{\KlebanovMS}.  

If we were to allow ourselves to take $Q \to \infty$, as in \refs{\SchmidhuberBV,\CooperVG},
then $\gamma Q \to 2$, and \firstordertwo\ becomes the standard renormalization
group equation.  Although this is a semiclassical limit of Liouville theory,
it is unclear to us whether or not terms higher order in $Q$ change this answer dramatically.
Nonetheless, since the answer makes sense, we offer the following comments.
First, because we have turned off the ``wrong branch'' solutions,
$\phi$-derivatives of $u(\phi)$ are order $1/Q$.  Secondly,
in the gravitational dressing formulae for $c < 1$ minimal models
coupled to gravity, $Q$ is order one and one might also expect corrections. 
Nonetheless the gravitational
dressing formulae which can be derived within our approximations are consistent with
matrix model computations (\cf\ \refs{\GinspargIS}\ for a review and references.)
Perhaps there is a scheme in which these formulae are exact.

\subsec{D-brane decay}

We conclude with a discussion of the well-studied case of
D-brane decay via open string tachyons,
and relate it to the picture developed here.  More generally, it should be little trouble
to adopt our picture to boundary perturbations.

Consider a $c=25$ conformal field theory times a timelike scalar
field $\phi$.  This scalar could be the Liouville mode of 2d gravity in conformal gauge.  
Sen \refs{\SenNU}\ has shown that the following boundary interaction
\eqn\specialboundaryint{
	\delta S = \lambda \oint \cosh \phi \equiv \oint V(\phi)
}
is an exactly marginal boundary operator describing the decay of a D-brane
at $\phi = 0$. The point is that the potential on the world-sheet gives an
energetic cost to boundaries supported away from $V = 0$ \refs{\HarveyNA}.

This coupling follows naturally from the formalism above:
in particular it is a solution to the following equations:
\eqn\tachyonequation{
	\ddot{u} + \bar\beta^u = \ddot u - u = 0
}
which we take to be the boundary analog of the tachyon beta function equations
of this system.  

The general solution to \tachyonequation\ is
\eqn\boundarymarginal{
	u(\phi) = a e^{\phi} + b e^{-\phi}
}
Different values of $a,b$ correspond to different tachyon profiles in time.
The solution $b = 0$ therefore describes the decay of the open string tachyon.  $a = 0$
describes the time-reversed version of this decay.  $a = b$ describes
a tachyon pulse arranged such that the tachyon begins and
ends in the closed string background, and at $\phi = 0$ is in the open string
background---this is conjectured to be an ``S-brane" \refs{\GutperleAI}.

A lesson of this exercise is that in open string theory,
both branches in \branches\ have a potentially physically 
sensible interpretation, so we should keep them. It seems sensible to do the
same in closed string theory as well.

Another point is that 
there are qualitative differences between the case of open and closed-string tachyon
decay.  In the closed string case, for tachyons which are not localized, we
cannot decouple closed string radiation.  Rather, as in the example discussed in \S4.2,
the tachyon can settle into the new minimum of the effective potential, 
transferring the potential energy released into the slope of the linear dilaton.

\newsec{Conclusions}

\subsec{A speculation}

Polyakov, in section 9 of  \refs{\PolyakovTP}, 
proposes a Hamilton-Jacobi formulation of the renormalization group equations 
coupled to two-dimensional gravity.  (Indeed, \refs{\PolyakovTP}\ was a 
major inspiration
for this present paper).  The initial conditions are the coupling constants $\bar u$ 
of a two-dimensional field theory $\CC_{\bar u}$.  
The Hamilton-Jacobi functional is the partition function $Z(\bar u)$ of this
field theory coupled to 2d gravity; $\bar u$ becomes the coupling at some particular
background value of the Liouville mode. (This picture has been discussed further in
\refs{\DharAI}, in which the authors have also noted the relation to the
``holographic RG" equations which emerge from the AdS/CFT correspondence.)

While there is much that we do not understand about \refs{\PolyakovTP,\DharAI},
they imply an interesting picture for the spacetime dynamics.    
A renormalization group trajectory is determined entirely by the couplings at some cutoff scale.  
When we couple the theory $\CC$ to an additional scalar, such as the Liouville
mode, by allowing the couplings in $\CC$ to depend on $\phi$,
this cutoff scale is exchanged for a background value $\bar\phi$ of the scalar,
and solutions to the conformal invariance conditions
are specified by $u(\bar\phi) = \bar u$ and $\dot u(\bar\phi)$.

In this work we have tried to interpret the equations of
motion for the tachyon as being some sort of
modified ``gravitational" renormalization group equations, 
which can be written in terms of the data of the underlying non-conformal
quantum field theory $\CC_{\bar u}$.   Can the additional initial conditions also 
be given an interpretation which is more intrinsic to the underlying field theory?

In a Hamiltonian formulation, one specifies initial coordinates and momentum.
If the initial coordinates in the Hamilton-Jacobi formulation are the couplings $\bar u$
of the underlying 2d field theory,  then the initial momenta are:
\eqn\hjmom{
	{\bar p}_a = \frac{\p}{\p\bar u^a} Z(\bar u) = 
	- \int d^2 z \left\langle \frac{\p u^b(\phi)}{\p \bar u^a}\CO_b(z) \right\rangle
}
Our speculation is that to specify trajectories of gravitationally dressed RG flows, 
one would specify both the couplings to and the vevs of $\CO$ in $\CC_{\bar u}$.
  
This is almost identical to the dual interpretation of the boundary conditions
on bulk fields in the AdS/CFT correspondence \refs{\BalasubramanianSN,\BalasubramanianDE}.
While the analogy is not precise, it is worth exploring. Studying the
Hamilton-Jacobi version \refs{\deBoerXF}\ of the
holographic renormalization group  equations \refs{\BalasubramanianJD}\
for a general class of backgrounds may provide some insight as to
how to think about this extra data in the context of renormalization
group flows in field theory.

\subsec{Additional questions}

There are many questions that remain.  For
example, we have worked with a highly restricted class of backgrounds.
We have set $G_{\phi\phi} = -1$,
and we have set couplings  of the form $u^a \p\phi \CO_a$ and
$\Phi^a \CO_a R^{(2)}(\phi)$ to zero.  Furthermore, our description
is not covariant from the spacetime point of view.  It would be useful to further explore spacetime
gauge transformations in this framework.  This may be an important part
of understanding the Hamilton-Jacobi formulation discussed above.

\bigskip
\centerline{\bf Acknowledgements}

We would like to thank Eric D'Hoker for collaboration and useful discussions on 
parts of this project, and Barton Zwiebach for helpful feedback on the manuscript.  We would like to thank Tom Banks, Stanley Deser, Jacques 
Distler, Ken Intriligator, Jesper Jacobsen, Martin Kruczenski, Jos\'e Latorre, 
Emil Martinec, John McGreevy, Shiraz Minwalla, Ronen Plesser, Howard 
Schnitzer, Stephen Shenker, Gary Shiu, Eva Silverstein, David Tong, 
Mark van Raamsdonk, and Barton Zwiebach for helpful conversations 
and comments. This paper is based in part on work done during the 2005 
String Theory  workshop at the Benasque Center for Science, which we would 
like to thank for hospitality. A.L. would also like to thank the Center for Geometry 
and Theoretical Physics at Duke University, the Fields Institute for Research in 
Mathematical Sciences at the University of Toronto, and the Perimeter Institute 
for Theoretical Physics for their hospitality during various parts of this project. 
The research of D.Z.F. is supported by the NSF grant PHY-00-96515. The 
research of M.H. is supported by a Pappalardo Fellowship and by the U.S. 
Department of Energy through cooperative research agreement 
DF-FC02-94ER40818. The research of A.L. is supported in part by 
NSF grant PHY-0331516, by DOE Grant No. DE-FG02-92ER40706, 
and by a DOE Outstanding Junior Investigator award.

\appendix{A}{Derivation of the renormalization group equation}

Our starting point is an ultraviolet fixed point described by a nontrivial
2d conformal field theory. These theories may be non-Gausssian -- the examples
we have in mind are the $c < 1$ unitary minimal models \refs{\BelavinVU}.

If the CFT can be described by an action $S_{CFT}$, 
we are interested in perturbed theories of the form
\eqn\pertcft{
	S = S_{CFT} + \int d^2 z \sum_A \eps^{\Delta_A - 2} u^A \CO_A(z) = S_{CFT} + \delta S\ .
}
where $u^a$ is a dimensionless coupling, $\Delta_A$ is the dimension of
$\CO_A$ in the unperturbed CFT, and $\eps$ is a length scale that we
will identify below with the cutoff.  If a Lagrangian description of the UV theory does not 
exist, then we can describe the perturbed theory via the correlation functions:
\eqn\corrpert{
	C_n(x_1,\ldots,x_n) = \langle \CO_1(x_1)\ldots \CO_n(x_n)\rangle_{u} \equiv
	\langle \CO_1(x_1)\ldots \CO_n(x_n) e^{ - \delta S}\rangle_{CFT}
}
The correlation function on the right hand side is meant to be evaluated in the CFT.

When the perturbing operators are not exactly marginal,
the theory \pertcft\ will run under renormalization group transformations.
We will review the derivation of the renormalization group equations for $\Gamma_n$
given by Zamolodchikov \refs{\ZamolodchikovTI}, albeit with slightly different
notation.  These equations describe the response of $C_n$ to
a rescaling of the arguments $x_k$.  We will rewrite them in a form which
describes the response of $C_n$ to a rescaling of the cutoff.

The Ward identity for scale transformations is:
\eqn\confwi{
	\left(\sum_k x_k\cdot\frac{\p}{\p x_k} + \hat{D}_k \right)
		\langle \CO_{A_1}(x_1)\ldots \CO_{A_n}(x_n)\rangle = 
		- \int d^2 z \langle \CO_{A_1}(x_1)\ldots\CO_{A_n}(x_n) \Theta(z) \rangle
}
where 
\eqn\sttrace{
	\Theta(z) = - \beta_w^A \CO_A \eps^{\Delta_A - 2}
}
is the trace of the stress tensor.\foot{The explicit factor of $\eps^{\Delta_A - 2}$ guarantees that
the stress tensor has engineering dimension $2$ near the UV fixed point. $\beta^A_w$ is
the dimensionless beta function for the dimensionless coupling $u^A$.}
The operator $\hat{D}$ is the dilatation operator acting on $\CO$:
\eqn\diloper{
	\CO_A(x + \lambda x) = \CO_A(x) + \lambda \hat{D} \CO_A + O(\lambda^2)
}
The ``beta functions" $\beta_w^A$ are here defined as the coefficients of the 
Weyl anomaly.  We have defined the sign of the beta function to correspond to 
particle physics conventions (so that the beta function is negative for asymptotically 
free theories).

The right hand side of \confwi\ can be rewritten in terms of a derivative of the 
correlation function with respect to $u$ as follows.
Derivatives of $\Gamma_n$ with respect to the couplings $u$ can be broken up 
into two pieces.  One comes from the change of the basis of operators $\Phi$ 
as a function of the couplings, while the second comes from the change in the action:
\eqn\couplingder{
\eqalign{
	\frac{\p}{\p u^C} \langle \CO_{A_1}(x_1)\ldots \CO_{A_n}(x_n) \rangle & =
	\sum_{k=1}^n \langle \CO_{A_1}\ldots B_c\CO_{a_k}(x_k)\ldots
	\CO_{A_n}(x_n)\rangle \cr
	& - \int d^2z \eps^{\Delta_c - 2}\langle \CO_{A_1}(x_1)\ldots\CO_{A_n}(x_n)
		\CO_C(z)\rangle
}}
If we contract \couplingder\ with $\beta_w^C$ and equate the terms with $\Theta$ in \couplingder\ and \confwi, we find that:
\eqn\rgequationpositionder{
	\left[ \left(\sum_k x_k \cdot \frac{\p}{\p x_k} + \Gamma_k \right)  
		+\beta_w^C \frac{\p}{\p u^C}\right]
		\langle \CO_{A_1}(x_1) \ldots \CO_{A_n}(x_n) \rangle = 0\ ,
}
where
\eqn\anomdim{
	\Gamma \CO_A(x) = \left(\hat{D}  - \beta_w^C\p_C\right) \CO_A(x) = \left( \delta^C_A \Delta_A
	- \gamma^C_A\right) \CO_C(x)
}
defines the anomalous dimension operator $\gamma$.
Note that the definition of $\gamma$ differs from that in \refs{\ZamolodchikovTI} 
by a sign and an additional factor of $\Delta_A$. Here it
is the deviation of the dimensions of $\CO_A$ from their values at the UV fixed point.

We can turn \rgequationpositionder\ into a differential equation in terms of the cutoff. 
If the only dimensionful parameter in the theory is $\eps$, then:
\eqn\correlationscaling{
	\Gamma_n(x_1,\ldots x_n) = \langle \CO_{A_1}(x_1)\ldots\CO_{A_n}(x_n)\rangle
	= \eps^{- \sum_k \Delta_k} F\left(\frac{\eps}{|x_i - x_j|}\right)\ ,
}
Therefore,
\eqn\derivativeswitch{
	\sum_k x_k \cdot \frac{\p}{\p x_k} \Gamma_n = - \left(\eps\frac{\p}{\p\eps} +
	\sum_K \Delta_K\right) \Gamma_n\ ,
}
so that we get the RG equation
\eqn\rgequation{
	\left( \eps\frac{\p}{\p \eps}  -\beta^C\p_C \right)
		\langle \CO_{A_1}(x_1)\ldots\CO_{A_n}(x_n)\rangle
		+ \sum_k \gamma_{A_k}^{C_k}
		\langle\CO_{A_1}(x_1)\ldots\CO_{C_k}(x_k)\ldots\CO_{A_n}(x_n)\rangle = 0
}

Zamolodchikov has also shown \refs{\ZamolodchikovGT,\ZamolodchikovTI}\ 
that 
\eqn\betatogamma{
	\p_A \beta^C = - \left(2 - \Delta_A\right) \delta^C_A - \gamma^C_A
}
where we have used our definition of $\gamma$ as the deviation 
from the dimension
of the operator at the UV fixed point.\foot{There is a typo in eg. (2.9) of \refs{\ZamolodchikovTI}\
which is crucial for this proof.  The rescaling of the unintegrated Euclidean 
Hamiltonian should read
$$H(x)\to H(x)(1+dt) - dt\Theta\ .$$
This can be checked by studying the case of a scalar field.  Note also that
there are various factor-of-two differences between the conventions in
\refs{\ZamolodchikovTI}\ and those of this work.}  Eq. \betatogamma\ is also consistent with
$\Theta$ having dimension exactly $2$.

\appendix{B}{The Zamolodchikov metric at order $u$ in the Wilsonian scheme}

Using Eq. \couplingder, we find that the derivative of the two-point function
with respect to the coupling is:
\eqn\cfderiv{
\eqalign{
	\frac{d}{d u^c} \langle \CO_a(x) \CO_b(y)\rangle|_{u=0} &=
	\frac{\p}{\p u^c} \langle \CO_a(x) \CO_b(y)\rangle|_{u=0}
	- \Lambda^k_{a,c} \langle \CO_k(x)\CO_b(y) \rangle  -
	\Lambda^k_{b,c} \langle \CO_a(x) \CO_k(y) \rangle\cr
	& = - \int d^2z \eps^{\Delta_c - 2} \langle \CO_c(z) \CO_a(x) \CO_b(y) \rangle
	+ \Lambda^k_{a,c} \langle \CO_k(x)\CO_b(y) \rangle \cr
	& \ \ \ \ \ \ \ \ \ \ +
	\Lambda^k_{b,c} \langle \CO_a(x) \CO_k(y) \rangle\cr
	& = - \frac{\CF\left(\frac{\eps}{|x-y|}\right)_{abc}
	}{|x-y|^{\Delta_a + \Delta_b}}+ 
	\frac{\eps^{\Delta_k - \Delta_a}\Lambda_{a,c}^k g^{(0)}_{kb}}
		{|x-y|^{\Delta_k + \Delta_b}} + 
	\frac{\eps^{\Delta_k - \Delta_b}\Lambda_{b,c}^k g^{(0)}_{ak}}
		{|x-y|^{\Delta_a + \Delta_k}}
}}
Here 
$$ B_c \CO_a = \eps^{\Delta_b - \Delta_a}\Lambda^b_{a,c}\CO_b\ , $$
where the factors of $\eps$ render $\Lambda$ dimensionless, and 
%
\eqn\cftzam{
	\langle \CO_a(x) \CO_b(y)\rangle|_{u=0} = \frac{g^{(0)}_{ab}}{|x-y|^{\Delta_a + \Delta_b}}\ .
}
In other words, $g_{ab}^{(0)}$ is the Zamolodchikov metric at the fixed point. 
Furthermore, we define: $\Lambda_{ab,c} = \Lambda^k_{b,c} g^{(0)}_{ka}$

The integrated three-point function is defined as:
\eqn\threepoint{
	\int d^2 z \eps^{\Delta_c - 2} \langle \CO_a(x)\CO_b(y)\CO_c(z)\rangle = 
	\frac{1}{|x-y|^{\Delta_a + \Delta_b}} \CF\left(\frac{\eps}{|x-y|}\right)_{abc}\ .
}
Note that this will contain cutoff-dependent terms in general.

If we define the Zamolodchikov metric for $u\neq 0$ as
\eqn\perturbedzam{
	g_{ab}(u) = \eps^{\Delta_a + \Delta_b} \langle \CO_a(\eps)\CO_b(0)\rangle\ ,
}
then \cfderiv\ implies that $g$ will
vanish to linear order in the couplings if
\eqn\condition{
   \CF_{abc}(1) = \Lambda_{ab,c} + \Lambda_{ba,c}
}

To lowest order in $u$, we can use the RG equations to find $\Lambda_{ab,c}$.  
This is because 
at linear order the exponential $e^{-\int u \CO}$ does not need to be regulated: all
of the divergences come from the contraction of the leading term in this exponential with
the operator insertions in the correlator.  The partial derivative of the two-point function
with respect to the cutoff is:
\eqn\scalederivative{
   \left[ \eps\frac{\p}{\p\eps}  \langle \CO_a(x)\CO_b(y)\rangle\right]_{x=\eps,y=0}
   = \left[- \CF'(1)_{abc} + \left(\Delta_b - \Delta_a\right)
   	\Lambda_{ab,c} + \left(\Delta_a - \Delta_b\right)\Lambda_{ba,c}\right] u^c
	\eps^{-\Delta_a-\Delta_b}
}
The $\beta^a\p_a$ term gives us, using \cfderiv,
\eqn\betaterm{
\eqalign{
   \beta^c \frac{\p}{\p u^c} \langle \CO_a(x) \CO_b(y) \rangle |_{x=\eps,y=0}
   & = - \beta^c\left( - \CF(1)_{abc} + \Lambda_{ab,c} + \Lambda_{ba,c}\right)
   \eps^{-\Delta_a - \Delta_b} \cr
   & = \left(2 - \Delta_c\right) \left[ - \CF(1)_{abc} + \Lambda_{ab,c} + \Lambda_{ba,c}\right]
   u^c \eps^{-\Delta_a - \Delta_b}
}}
where $c$ is summed over.

Assuming that $\gamma$ begins at order $u$, we can add all of the $O(u)$ terms
together in \rgequation\ to find:
\eqn\linearrgequation{
\eqalign{
   -\CF'(1)_{abc}u^c & - (2 - \Delta_c)  \CF(1)_{abc}u^c \cr
   &+  (2 - \Delta_{ac,b})\Lambda_{ab,c} +
   (2 - \Delta_{bc,a})\Lambda_{ba,c} + \gamma^k_a g^{(0)}_{kb}
   + \gamma^k_b g^{(0)}_{ak} = 0
}}
where $\Delta_{ab,c} = \Delta_a + \Delta_b - \Delta_c$.

If we can compute $\CF$, then $\Lambda$ is determined to the extent that
we need to by \linearrgequation. If we do not impose a cutoff,
the three-point function \threepoint\ is determined by conformal invariance:
\eqn\threepointconformal{
	\langle \CO_a(x) \CO_b(y)\CO_c(z)\rangle = \frac{C_{abc}}
		{|x-y|^{\Delta_{ab,c}}|x-z|^{\Delta_{ac,b}}|y-z|^{\Delta_{bc,a}}}
}
If $\Delta_a = 2 - \delta_a$, $\delta_a \ll 1$, then these integrals will not be
IR divergent, but they potentially have UV divergences.  However,
if we conttinue $\delta_a$ to a region where these divergences
are absent, then the integral over $z$ can be done without
imposing an additional cutoff \refs{\ZamolodchikovTI}:
\eqn\integratedthreepoint{
\eqalign{
	& \int d^2 z \eps^{\Delta_c - 2}\langle \CO_a(x)\CO_b(y)\CO_c(z)\rangle\cr
		& = \pi C_{abc} \frac{\Gamma(\Delta_c - 1) 
		\Gamma\left(1 + \frac{\Delta_a - \Delta_b - \Delta_c}{2}\right) 
		\Gamma\left(1 + \frac{\Delta_b - \Delta_a - \Delta_c}{2}\right)}
		{\Gamma\left(\frac{\Delta_a + \Delta_c - \Delta_b}{2}\right)
		\Gamma\left(\frac{\Delta_b + \Delta_c - \Delta_a}{2}\right)
		\Gamma(2 - \Delta_c)}\frac{1}{|x-y|^{\Delta_a + \Delta_b + \Delta_c - 2}}\cr
		& = \frac{P_{abc}(\Delta_{a},\Delta_b,\Delta_c)}{|x-y|^{\Delta_a + \Delta_b}} 
		\left(\frac{\eps}{|x-y|}\right)^{\Delta_c - 2} = 
	\frac{{\cal I}_{abc}(|x-y|)}{|x-y|^{\Delta_a + \Delta_b}}.
}}
(Note that the definition of $\Delta$ in this paper is twice that used in \refs{\ZamolodchikovTI}.)
However, for most of this paper we are adopting a scheme where the OPEs of operators
in the action are cut off regardless of whether the OPEs lead to singularities.
As in \ZamolodchikovTI, the difference will appear as a renormalization of the operators:
\eqn\realthreepoint{
	\int d^2 z \eps^{\Delta_c - 2}\langle \CO_a(x)\CO_b(y)\CO_c(z)\rangle =
	\frac{\CI_{abc}}{|x-y|^{\Delta_a + \Delta_b}} + \langle (\hat b_c \CO_a)(x)\CO_b(y) \rangle +
	\langle \CO_a(x) (\hat b_c \CO_b)(y)\rangle
}
The cutoff dependence is determined by dimensional analysis:
\eqn\renormcutoff{
	\langle (\hat b_c \CO_a)(x) \CO_b(y)\ldots\rangle
	= \eps^{\Delta_k - \Delta_a} B_{ac}^k \langle \CO_k(x) \CO_b(y)\ldots\rangle
}

We can rewite the numerator in the last line of \cfderiv\ as
\eqn\numerator{
	\frac{d}{d u^c}\langle \CO_a(x)\CO_b(y)\rangle
	= \frac{- \CI_{abc} + \hat\Lambda_{ab,c} + \hat\Lambda_{ba,c}}{|x-y|^{\Delta_a + \Delta_b}}
}
where 
\eqn\lambdashift{
\eqalign{
	\hat \Lambda_{ab,c} & = \Lambda_{ab,c} - B_{ab,c} \cr
	B_{ab,c} & = B_{ac}^k g_{kb}
}}
Similarly, we can rewrite \linearrgequation:
\eqn\linearrgequation{
\eqalign{
   -\CI'(1)_{abc}u^c & - (2 - \Delta_c)  \CI(1)_{abc}u^c \cr
   &+  (2 - \Delta_{ac,b})\hat \Lambda_{ab,c}u^c +
   (2 - \Delta_{bc,a})\hat \Lambda_{ba,c}u^c + \gamma^k_a g^{(0)}_{kb}
   + \gamma^k_b g^{(0)}_{ak} = 0.
}}

Now the cutoff-independent part $I_{abc}$ of the three-point function can be written as:
\eqn\scalingfunction{
	\CI_{abc} = \pi C_{abc} \frac{\Gamma(\Delta_c - 1) 
		\Gamma\left(1 + \frac{\Delta_a - \Delta_b - \Delta_c}{2}\right) 
		\Gamma\left(1 + \frac{\Delta_b - \Delta_a - \Delta_c}{2}\right)}
		{\Gamma\left(\frac{\Delta_a + \Delta_c - \Delta_b}{2}\right)
		\Gamma\left(\frac{\Delta_b + \Delta_c - \Delta_a}{2}\right)
		\Gamma(2 - \Delta_c)} \left(\frac{\eps}{|x-y|}\right)^{\Delta_c - 2}
}
We can expand this ratio of gamma functions to leading order in $\Delta - 2$:
\eqn\expandgamma{
	P_{abc} = C_{abc} \frac{4\pi(2 - \Delta_c)}{(2-\Delta_{ab,c})
		(2-\Delta_{bc,a})}\left(1 + O\left((2 - \Delta)^3\right)\right)
}

Eq. \scalingfunction\ implies that $\CI'(1)_{abc} = (\Delta_c - 2)\CI(1)_{abc}$.
Therefore the first two terms of \linearrgequation\ cancel, leaving:
\eqn\simplifiedequation{
	(2 - \Delta_{ac,b})\hat \Lambda_{ab,c} +
   (2 - \Delta_{bc,a})\hat \Lambda_{ba,c} + \gamma^k_a g^{(0)}_{kb}
   + \gamma^k_b g^{(0)}_{ak} = 0
}
To quadratic order in perturbation theory, based on \betatogamma, $\gamma^c_a=-4\pi C^c_{ab}u^b$. The equation \simplifiedequation\ is solved if we set:
\eqn\gammatolambda{
	\hat \Lambda_{ab,c} = \frac{2\pi C_{abc}}{2 - \Delta_{ac,b}}.
}
This means that \cfderiv\ is zero, and the Zamolodchikov metric has no linear term, if
\eqn\zamcondition{
	\CI(1)_{abc} = \frac{4\pi C_{abc} (2 - \Delta_c)}{\left(2-\Delta_{ac,b}\right)
	\left(2 - \Delta_{bc,a}\right)}.
}
This follows from the leading form \expandgamma.

\appendix{C}{The beta functions at second order in derivatives}

In \S3.2, we computed the beta function of \coupledtheory\ 
to lowest order in a perturbation series in $\phi$-derivatives
and in $\delta_a = 2 - \Delta_a$.  In this appendix we will
exhibit more explicitly the expansion in $\delta$.  Although the
calculation simplifies at the order in which we are working,
we will show where higher order terms could appear.

The major issue that arises is the proper treatment of infrared divergences.
These divergences plague us for two reasons.  First,
the derivative expansion is one about a massless scalar field
theory in two dimensions.  In this theory the scalar fields themselves are
not well defined as quantum operators, due to infrared divergences.
Secondly, the perturbations in \coupledtheory\ are naively relevant, at order $\delta_a$.
Such perturbations generically lead to infrared divergences.
These divergences signal that matrix elements of operators
are nonanalytic in the coupling constants and in derivatives with respect to $\phi$.
The ultraviolet properties of the theory should be free of these divergences.
Let us discuss our treatment of each of these issues in turn.

The definition of the scalar field $\xi$ haunts us for the following reason.
In \actionexpand, we can treat $\dot{u},\ddot{u},\ldots$ as couplings to 
composite operators made up from $\CO_a$ and powers of $\xi$.
As we stated above, $\xi$ are not good quantum operators in two dimensions,
due to the large infrared divergences of massless fields in two dimensions.
However, because we are interested in the ultraviolet properties of our model, 
we should be able to regulate these divergences, taking
care to insure that the beta functions do not depend on the IR regulator.
Following \refs{\AmitAB}, we can regulate the divergences by adding
a small mass $R^{-1} \equiv m$ to $\xi$, leading to the propagator
\eqn\mpropagator{
	\langle \xi(z)\xi(0)\rangle = 2 K_0 \left( m (|z|^2 + \eps^2)\right)
}
if $|z|,\eps \ll m^{-1}$, then this can be approximated by
\eqn\propagator{
	\langle \xi(z) \xi(0) \rangle = \ln \left( \frac{|z|^2 + \eps^2}{R^2}\right)
}
which is the propagator we were using in \S3.2.   In practice we will use this
latter form of the propagator, as we will be needing to impose additional
infrared cutoffs as well.  Note that the sign of the propagator is that of a
scalar field with the ``wrong sign" kinetic term, corresponding to a timelike
target space direction.

In the limit $m \to 0$, we should think of the beta functions for these composite operators
as follows.  The theory \coupledtheory\ will contain divergences multiplying
operators of the form $f^a(\phi) \CO_a$.  
One cancels these divergences with cutoff-dependent counterterms,
of the form $\delta u^a_{\eps} \CO_a$, according to one's renormalization scheme.
These should be good operators~-- for example, $\delta u$ may take the
form of an exponential of $\phi$.  

The beta functionals $\beta^a(u(\phi))$ are derivatives of the counterterms 
$u_{\eps}$ with respect to the
cutoff $\eps$.  One may expand the original divergent terms $F^a(\phi)\CO_a$
as well as the counterterms:
\eqn\counterterms{
\eqalign{
	f^a(\phi) \CO_a & = \left( f^a(\bar\phi) + \xi \dot f^a (\bar\phi) + :\xi^2: \ddot
		f^a(\bar\phi) + \ldots \right) \CO_a \cr
	u^a_{\eps}(\phi)\CO_a  & = \left(u^a_{\eps}(\bar\phi) 
		+ \dot u^a_{\eps}(\bar\phi) \xi + \half \ddot u_{\eps}^a(\bar\phi):\xi^2: 
		+ \ldots\right) \CO_a
}}
The beta functions $\beta^{a,k}$ for the ``operators" $\xi^k \CO_a$ are the 
logarithmic derivatives of
$\p_{\phi}^k \delta u^a$ with respect to the scale $\eps$. 

In general, we should be able to expand about any point $\bar\phi$, to find the
beta functions for $u^a(\bar\phi)$ as a function of $\bar\phi$.  This means that
$\beta^{a,k}$ should satisfy consistency conditions \refs{\FriedanJM}, {\it e.g.}\
\eqn\consist{
	\beta^{a,1} = - \p_{\phi} \beta^a(u(\phi)) = - \dot{u}^b\p_b \beta^a
}
We will keep $\beta^{a,k}$ explicit, however, and find that \consist\ is indeed
satisfied at the order in which we are working.

The next set of infrared problems will arise because we are studying 
perturbations by operators
$u^a(\phi)\CO_a$\ which are naively relevant.  However, as we have discussed
in \S3.2, we are performing a double expansion in $\phi$-derivatives and in $\delta_a = 2 - \Delta_a$,
with $\delta_a$ the same order as two derivatives.  We
implement this by expanding the OPE coefficients:
\eqn\opeexpand{
\eqalign{
	\CO_a(z) \CO_b(w) & \sim \frac{\CC_{ab}^c\left(u,\frac{\eps^2}{|z|^2 + \eps^2}\right)} 
		{\left(|z-w|^2 + \eps^2\right)^{1 - \delta_{ab,c}/2}}\CO_c(\bar z)\cr
		& \sim \frac{\CC_{ab}^c(u,1) \CO_c(\bar z) }
		{|z-w|^2 + \eps^2} \left(1
		+ O(\delta) + \ldots\right)
}}
where $\delta_{ab,c} = \delta_a + \delta_b - \delta_c$, and $\bar z = \half(z+w)$.  
The $O(\delta)$ terms will include logarithms of $|z|^2+\eps^2$, while
the additional terms will include terms analytic in $\frac{\eps^2}{|z|^2 + \eps^2}$.

Because $\CO_a$ are treated as marginal, we can write the OPEs at lowest
order in $\delta$ as (\cf\ \refs{\KutasovXB}):
\eqn\opeexpand{
	\CO_a(z) \CO_b(w)  \sim \frac{\CC_{ab}^c(u)}{|z-w|^2 + \eps^2}\CO_c(\bar z)
		- \Gamma^c_{ab}(u)\CO_c(\bar z) \delta^2(z-w) + \ldots
}
The second term corresponds to a contact term, and
the reason for the sign in front of it will become clearer below.
The first term on the left hand side is known to lead to anomalous dimensions
proportional to $\CC_{bc}^a$.  For our assumptions to be self-consistent,
$\CC$ must also be order $\delta$.

For the theory \coupledtheory, we can write the RG equation for the partition
function as:
\eqn\coupledRG{
	\eps\frac{\p}{\p\eps} Z^{\eps} 
		-\beta^a(u(\bar{\phi}))\frac{\p}{\p u^a(\bar{\phi})} Z^{\eps} -
		\sum_{k=1}^{\infty} \beta^{a,k}\frac{\p}{\p u^{(k)}} Z^{\eps} = 0\ ,
}
where
\eqn\fielderiv{
	u^{(k)} = \frac{\p^k}{\p\bar{\phi}^k} u(\bar{\phi})
}

We would like to compute $\beta$ to second order in $\delta^a$ and in $\phi$-derivatives,
assuming that we know the beta functions and anomalous dimensions for \constanttheory.
The expansion of the partition function to second order in $\xi$ is:
\eqn\pertexpansion{
\eqalign{
	Z & = \langle e^{-\sum_a \int d^2 z \eps^{\Delta_a - 2} u^a (\bar{\phi}) \CO_a(z) }\cr
	& \times \left( 1 + \int d^2z \eps^{\Delta_a - 2} \dot{u}^a(\bphi) \xi(z)\CO_a(z)
	- \half \int d^2 z \eps^{\Delta_a - 2} \ddot{u}^a (\bphi)\xi^2(z) \CO_a(z) \right. \cr
	&+ \left. \half \int\int d^2 zd^2 w \eps^{\Delta_a + \Delta_b - 4}
	\dot{u}^a(\bphi)\dot{u}^b(\bphi) \xi(z)\xi(w)
		\CO_a(z)\CO_b(w)
	 + \ldots \right)\rangle_{\eps,u(\bphi)}
}}
Here the expectation values are all taken in the tensor product of
\constanttheory\ with the free scalar theory for $\xi$. 

We will use the regulator \cptope\ for OPEs of the operators $\CO_a$,
and \propagator\ for the OPEs of $\xi$.
Note that using \propagator\ to compute
composite operators of the form $e^{ik\phi}$ will lead to OPEs of these operators
of the form \cptope.

There are divergences in \pertexpansion\ that arise from defining the exponential.
These divergences, inserted into
\rgequation, give the beta functions 
$\bar{\beta}^a(u(\bar{\phi}) = - \eps\frac{\p}{\p\eps} u^a(\bar{\phi})$ 
of the theory \constanttheory.  We are taking these as given. 
The new divergences that appear to this order come from:

\noindent\item{{\bf 1.}} The OPE singularities of the fields within the
parentheses of \pertexpansion\ with each other, using \cptope\ and \propagator.

\noindent\item{{\bf 2.}}  The OPE singularities of the fields in parentheses of
\pertexpansion\ with the exponential.

We first expand perform the contractions in \pertexpansion\ to write $Z$ at lowest order
in $\delta$ and $\phi$ derivatives as:
\eqn\newdivergences{
\eqalign{
	Z^{\eps} &= \langle e^{-\sum_a \int d^2 z u^a (\bar{\phi}) \CO_a(z)}
	\ \left( 
	1 - \int d^2 z  \dot{u}^a \xi^z \CO_a(z) \right. \cr
	&\left. -\half \int d^2 z \ddot{u}^a(\bar{\phi})\left(\ln\frac{\eps^2}{R^2}\right)
	\CO_a(z)
	- \half \int d^2 z \ddot{u}^a :\xi^2: \CO_a(z) \right. \cr
	&\left.+ \half \int d^2zd^2y
	\frac{\CC^a_{bc}
	\dot{u}^b\dot{u}^c}{\left(|y|^2+\eps^2\right)}
	\ln\left(\frac{|y|^2 + \eps^2}{R^2}\right) \CO_a(z)\right.\cr
	 &\left.+ \half \int d^2z d^2y\ 
	\frac{\CC^a_{bc}
	\dot{u}^b\dot{u}^c}{\left(|y|^2+\eps^2\right)}
	:\xi^2: \CO_a(z)+ \right.\cr
	& \left. + \half \int d^2 z \Gamma^a_{bc}(u) \dot u^b \dot u^c\ln \left(\frac{\eps^2}{R^2}\right)
	\CO_a(z) + \half \int d^2 z \Gamma^a_{bc} \dot u^b \dot u^c :\xi^2:\CO_a(z)
	\ldots\right)\rangle
}}
The normal ordering symbols correspond to the composite operators
with the contractions explictly subtracted. 

Now we would like to perform the integrals $d^2 y$ in the third and fourth lines.
In fact, since $\CC$ is order $O(\delta)$, these are automatically
higher order in $\delta$.  Furthermore, if we include additional higher-order
terms from the expansion in $\delta$, these will give us integrals
of the form
$$ \int d^2 z \frac{\delta^k \left(\ln\frac{|z|^2 + \eps^2}{R^2}\right)^k}{|z|^2 + \eps^2}
\sim - \delta^k \left(\ln\frac{\eps^2}{R^2}\right)^{k+1} $$
where $k \geq 1$. These ``nonlocal" terms
correspond to overlapping divergences and will drop out of the computation
of the beta functions.  Such terms appear at higher order than the order
we are working in.  We have checked that at order
$\delta\p_{\phi}^2$, and at second order in $u$, such nonlocal terms also cancel.

The resulting ``local" terms in the partition function are:
\eqn\integratedZ{
\eqalign{
	Z^{\eps} & = \langle e^{-\sum_a \int d^2 z  u^a (\bar{\phi}) \CO_a(z)}
	\ \left( 
	1 - \int d^2 z  \dot{u}^a \xi^z \CO_a(z) \right. \cr
	&\left. -\half \int d^2 z\ddot{u}^a(\bar{\phi})\left(\ln\frac{\eps^2}{R^2}\right)
	\CO_a(z)
	- \half \int d^2 z \ddot{u}^a :\xi^2: \CO_a(z) \right. \cr
	&\left. + \half \int d^2 z \Gamma^a_{bc}(u) \dot u^b \dot u^c\ln \left(\frac{\eps^2}{R^2}\right)
	\CO_a(z) + \half \int d^2 z \Gamma^a_{bc} \dot u^b \dot u^c :\xi^2:\CO_a(z)
	\ldots\right)\rangle
}}

The first step in our calculation is to compute the 
explicit derivative $\eps\p_{\eps}\Gamma$ in \coupledRG.  This will act both on the
explicit factors on $\eps$ in \newdivergences, and on the expectation values
$\langle\CO\rangle$ in \constanttheory.  The latter are determined
by the RG equation \rgequation\ for \constanttheory:
\eqn\intermediatecs{
\eqalign{
	0 & = \eps\frac{\p}{\p\eps} \langle \CO_a(z)\rangle - 
	 \bar{\beta}^b \langle \int d^2 y \CO_b(y) \CO_a(z)\rangle
	 - \bar{\beta}^b B_{a,b}^c \langle \CO_c \rangle
		+ \bar{\gamma}^c_a \langle \CO_c(y) \rangle\cr
	& = \eps\frac{\p}{\p\eps} \langle \CO_a(z) \rangle
		 - \bar{\beta}^b \int d^2 y \frac{\CC^c_{ba}}
		{\left(|y-z|^2+\eps^2\right)^{1 - \delta_{abc}/2}} \langle \CO_c\left(
			\frac{y+z}{2}\right)\rangle
		+ \tilde{\gamma}^c_a \langle \CO_c(y) \rangle
}}
Here we have defined
\eqn\shiftad{
	\tilde \gamma^a_b = \bar\gamma^a_b - \bar \beta^c \Lambda_{b,c}^a
}
where $\bar\gamma$, $\bar\beta$ are the anomalous dimensions and beta functions
for \constanttheory, and $B_a \CO_b = \Lambda^k_{b,a} \CO_k$, with $B$ defined
in \couplingder. 

Keeping only terms to second order in derivatives {\it or}\ to first order in $\delta$, we find:
\eqn\newscalederiv{
\eqalign{
	& \eps\frac{\p}{\p\eps} Z^{\eps} \cr
	& =
	- \int d^2 z \left(\bar{\beta}^a(u(\bar{\phi}))
	- \ddot{u}^a(\bar{\phi}) -
	\Gamma^a_{bc}\dot u^b \dot u^c \right)\langle \CO_a(z)\rangle_{\eps}\cr
		&+ \int d^2z \left(\delta_a\dot{u}^a - \dot{u}^d\tilde{\gamma}_d^a\right)
			\langle\xi\CO_a(z)\rangle_{\eps}\cr
		&+ \int d^2z d^2w \dot{u}^b\bar{\beta}^c
			\langle\xi\CO_b(z)\CO_c(w)\rangle_{\eps} + \ldots
}}
where we have dropped terms proportional to $\xi^k \CO$ for $k > 1$.

The ``local" terms in $\beta^\a\p_a$ are:
\eqn\betaderivative{
	\beta^a\frac{\p}{\p u^a} Z^{\eps}  = 
	\int d^2z  \beta^a(u) \CO_a(z)	
	-\int d^2zd^2w\ \dot{u}^b \beta^c\langle\xi\CO_b(z)\CO_c(w)\rangle
	+\ldots
}
The derivative $\beta^{(a,1)}\p_{\dot{u}^a}$ in \coupledRG\ gives us:
\eqn\betadotderivative{
	\beta^{a,1}\frac{\p}{\p\dot{u}^a}Z  =  \int d^2z\beta^{(a,1)}
	\langle\xi\CO_a(z)\rangle
}

Combining eqs. \newscalederiv,\betaderivative,\betadotderivative, we find that:
\eqn\finalbeta{
	\beta^a = - \ddot u^a - \Gamma^a_{bc} \dot u^b \dot u^c + \bar\beta^a
}
The beta function for $\dot{u}^a(\bar{\phi})$ is:
\eqn\derivbeta{
\eqalign{
	\beta^{(a,1)} & = - (2-\Delta_a)\dot{u}^a 
	+ \gamma^a_b\dot{u}^b
	- \int d^2w \frac{\CC_{bc}^a \eps^{\Delta_{bca}-2}}{\left(|w|^2+\eps^2\right)^{\Delta_{bca}/2}}
	\dot{u}^b\left(\beta^c -\bar{\beta}^c\right)\cr
	& = - (2-\Delta_a)\dot{u}^a - \gamma^a_b\dot{u}^b + {\hbox{higher derivative terms}}
}}
Using the relation \betatogamma, we find that \consist\ is satisfied to the order in which 
we are working.

Our final task is to interpret the contact term $\Gamma$.  As in \refs{\KutasovXB}, we claim
that it is a connection compatible with the Zamolodchikov metric in \constanttheory.  
We must modify the argument, since the operators are nearly rather than exactly marginal;
in our case the statement made in \refs{\KutasovXB}\ is only true to order $O(\delta_a)$.

The argument runs as follows.
The Zamolodchikov metric is defined for us as the two-point function
\eqn\zmetricdef{
	g_{ab} = \eps^{\Delta_a + \Delta_b} \langle \CO_a(\eps) \CO_b(0)\rangle_{u,\eps}
}
Now, let us take a derivative of this term with respect to the coupling $u$:
\eqn\linearderiv{
	\p_c g_{ab} = \Lambda_{a,c}^k \langle \CO_k(\eps) \CO_b(0) \rangle
	+ \Lambda_{b,c}^k \langle \CO_a(\eps) \CO_k(0)\rangle 
	- \int d^2 z \langle \CO_c(z)\CO_a(\eps)\CO_b(0) \rangle
}
To evaluate the second term, we require the OPE coefficients.  To leading order
in $\delta$, these are given by \opeexpand, so that
\eqn\derivagain{
\eqalign{
	\p_c g_{ab} & - \Gamma^k_{ac}g_{kb} - \Gamma^k_{bc} g_{ak}
	= \cr
	& \Lambda_{a,c}^k  \langle \CO_k(\eps)\CO_b(0)\rangle
	+ \Lambda_{b,c}^k  \langle \CO_a(\eps)\CO_b(0)\rangle\cr
	&
	- \int d^2 z 
	\frac{\CC_{abc}}{(|z-\eps|^2+\eps^2)(|z|^2 +\eps^2)(2\eps^2)}
}}
As we stated below \opeexpand, the final term is order $O(\delta)$.  Now,
at lowest order in $\delta$, $\CO_a$ are exactly marginal, as $\bar\beta$ are
by assumption $O(\delta)$ for general $u$.  For exactly marginal operators,
we may choose a basis such that $\Lambda = 0$.  Any obstruction to
this will appear at the order $\delta$ at which the theory fails to be conformal.
This is a particular choice of scheme in \constanttheory.
 
Therefore, to lowest order in $\delta$, 
\eqn\covderivagain{
	\nabla_c g_{ab} \equiv \p_c g_{ab} - \Gamma^k_{ac} g_{kb} - \Gamma^k_{bc}g_{ak}
		= 0
}
and so we can regard $\Gamma$ as the Christoffel connection for the metric $g$.

\listrefs
\end